\documentclass[3p,number,sort&compress]{elsarticle}

\journal{Annals of Physics}

\usepackage{epsfig,amsmath,amssymb,mathtools,amscd}
\usepackage[active]{srcltx}
\usepackage{dsfont}
\usepackage{subfigure}
\usepackage{enumerate}
\newtheorem{lemma}{Lemma}
\newtheorem{proposition}{Proposition}

\mathchardef\minus="002D 
\def\Proof{\medskip\par\noindent{\bf Proof. }}
\def\qed{$\,\blacksquare$\par}
\def\<{\langle}\def\>{\rangle}
 \def\ket#1{| #1 \rangle} 
\def\bra#1{\langle #1 |}
\def\ketbra#1#2{| #1 \rangle \langle#2 |}
\def\braket#1#2{\langle #1 | #2 \rangle} 

\def\v#1{\boldsymbol{\mathrm #1}} 
\def\Z{\mathbb Z} 
 
\def\Reals{\mathbb{R}}
 
\def\Hilb{\mathcal H}
\def\Neigh{\mathcal N} 

\def\L2{{\mathcal L}_2}

\def\s{\bar{s}}

\def\d#1 {\mathop{\!\! \mathrm{d}#1}\,}
\def\df#1#2 {\!\!\frac{\mathop{\mathrm{d}#1}}{#2}\,}

\newcommand{\Tr}{\mathop{\mathrm{Tr}}}

\begin{document}
\renewcommand{\arraystretch}{1.5}

\begin{frontmatter}

\title{Quantum Field as a quantum cellular automaton: the Dirac free evolution in one dimension}

\author{Alessandro Bisio}
\author{Giacomo Mauro D'Ariano}
\author{Alessandro Tosini}

\address{Dipartimento di Fisica dell'Universit\`a di Pavia, via Bassi 6, 27100 Pavia}
\address{Istituto Nazionale di Fisica Nucleare, Gruppo IV, via Bassi 6, 27100 Pavia}

\date{\today}

\begin{abstract}

We present a quantum cellular automaton model  in one space-dimension
which has the Dirac
  equation as emergent.
This model, a discrete-time and causal unitary  evolution of a lattice of quantum systems,
 is derived from the assumptions of  homogeneity, parity
  and time-reversal invariance.
  
  The comparison between the automaton and the Dirac evolutions is
  rigorously set as a discrimination problem between unitary
  channels. We derive an exact lower bound for the probability of
  error in the discrimination as an explicit function of the mass, the
  number and the momentum of the particles, and the duration of the
  evolution. Computing this bound with experimentally achievable
  values, we see that in that regime the QCA model cannot be discriminated
  from the usual Dirac evolution.

Finally, we show that the evolution of one-particle states with
narrow-band in momentum
can be efficiently simulated by a dispersive differential
  equation for any regime. This analysis allows for 
  a comparison with the dynamics of  wave-packets as it is described by the
  usual Dirac equation.

This paper is a first step in exploring the idea that quantum field
theory could be grounded on a more fundamental quantum
cellular automaton model and that physical dynamics could 
emerge from quantum information processing.
 In this framework, the discretization is a central ingredient and not only
  a tool for performing non-perturbative calculation as
  in lattice gauge theory. The automaton model, endowed with
  a precise notion of local observables and a full
  probabilistic interpretation, could lead to a coherent unification of an
  hypothetical discrete Planck scale with the usual Fermi scale of
  high-energy physics.
\end{abstract}


\end{frontmatter}

\section{Introduction}



The major problem of developing a quantum theory of gravity, whose
effects should become relevant at the Planck scale, seems to require a
deep reconsideration of the spacetime structure.  Recently alternative
models of spacetime are gathering increasing attention. We can cite
for example the loop quantum gravity model by Rovelli, Smolin and
Ashtekar \cite{rovelli1990loop} \cite{ashtekar1992weaving}
\cite{rovelli1995discreteness}, the causal sets approach of Bombelli
\emph{et al.}  \cite{bombelli1987space}, the noncommutative spacetime
of Connes \cite{connes1991particle}, the quantized spacetime of Snyder
\cite{snyder1947quantized}, the doubly-special relativity of Camelia
in \cite{amelino2002relativity}\cite{amelino2001planck} along with the
deformed special relativity models of Smolin and Magueijo in
\cite{magueijo2003generalized}. Some of these approaches are even
considered for experimental tests, see for example the recent
experiment proposals by Hogan \cite{hogan2010interferometers},
\cite{hogan2012covariant} and Brukner \cite{pikovski2011probing}.
Moreover, the finiteness of the entropy of a black hole
\cite{bekenstein1972black,hawking1975particle}, which implies
that the number of bits of information that can be stored is
finite, has led to the idea that space-time at the Planck scale could
be discrete and that the amount of information in a finite volume must
always be finite. 

In this work, following the ideas proposed in
Refs. \cite{darianovaxjo2010,darianovaxjo2011,darianopla,darianovaxjo2012,darianosaggiatore},
we assume that at the Planck scale physical dynamics occurs on a
discrete lattice and in discrete time steps. Considering for simplicity
the one-dimensional case, the lattice is a chain of sites equally
spaced with a period assumed to be equal to the Planck length
$\ell_P$, while a single time step is equivalent to a Planck time
$\tau_P$.  Each site $x$ corresponds to a quantum system whose
dynamics is described by a \emph{quantum cellular automaton} (QCA).
The QCA generalizes the notion of cellular automaton of von Neumann
\cite{neumann1966theory} to the quantum case, with cells of quantum
systems interacting with a finite number of nearest neighboring cells
via a unitary operator describing the single step evolution.

One of the first theoretical notion of QCA appeared in
Ref.~\cite{watrous1995one}, and later in
\cite{durr1996decision,durr1996decision-book} where it was referred to
as linear quantum cellular automata, while the notion of QCA as a mean
for simulating quantum physical systems originally appeared in
Refs.~\cite{boghosian1998quantum,boghosian1998simulating,love2005dirac}.
Since then the QCAs have been a quantum-computer-science object of
investigation with a rigorous formulation and relevant results about
their general structure \cite{schumacher2004reversible,
  arrighi2011unitarity, gross2012index}.  Moreover, in the field of
quantum information, particular attention is devoted to the so-called
\emph{quantum walks} (QWs) which describe the quantum evolution of one
particle moving on a discrete lattice and which correspond the one
particle sector of QCAs with linear evolution
\cite{grossing1988quantum,aharonov1993quantum,ambainis2001one}\footnote{Notice
  that in Ref.~\cite{grossing1988quantum} the word quantum cellular
  automaton appears for the first time. However, the model presented
  in the paper describe the one-particle evolution and is technically
  a QW.}.  This interest is motivated by the use of QWs in the design
of quantum algorithms: in Ref.~\cite{childs2003exponential} Childs et
al.  proved that QWs provide an exponential speedup for an oracular
problem and QWs are also known to provide polynomial speedups for many
relevant problems \cite{ambainis2007quantum, magniez2007quantum,
  farhi2007quantum}.

The idea of modeling the physical evolution at the Planck scale
on a discrete background first appeared
 in the work of 't Hooft \cite{t1990quantization}.
However, in his work the automaton is
classical, and it describes a deterministic discrete
theory underlying quantum theory. Then the idea of using
QWs for the simulation of Lorentz-covariant differential equations
appeared in the pioneering works of Succi and Benzi \cite{succi1993lattice},
Bialynicki-Birula \cite{bialynicki1994weyl} and in the  context of
lattice-gas simulations, especially in the works of Meyer 
\cite{meyer1996quantum} and Yepez
\cite{Yepez:2006p4406}.

It is important to stress that the approach we are proposing does not
aim to a  QCA-discretization of the known standard model of particle physics.
Indeed, we do not want to determine the QCA dynamics by mimicking the known dynamics of
Quantum fields but we propose to derive it  from principles of symmetry and
simplicity of the quantum algorithm. Clearly, because of the discreteness of this
framework, the usual continuous symmetries (like the Poincar\'e
invariance and the gauge symmetries) are no longer tenable and must be
replaced.  However, in the QCA model one can naturally require the
invariance of the dynamics under the discrete symmetries of the
lattice (like translation invariance, reflections and discrete
rotations).  In this work we consider a one dimensional QCA model
which is linear\footnote{Because of the linearity assumption one can
  regard this model as a second quantized version of a quantum walk}
and which has the minimal number of internal degrees of freedom for a
non-trivial evolution.  We then show that it is possible to
single out a class of unitarily equivalent QCA by imposing the symmetry
under discrete translations, parity and time reversal. Among the QCA
in this class, we then focus on the one whose expression
reproduces the Dirac equation in the Weyl representation as a finite
difference equation.

If the QCA model is a valid description of the microscopic dynamics, then
it must recover the usual phenomenology of quantum field
theory (QFT) at the energy scale of the current particle physics
experiments. This means that the physics of the QCA model and the one
of QFT must be the same as far as we restrict to quantum states that
cannot probe the discreteness of the underlying lattice. It is then
crucial to address a rigorous comparison between the QCA dynamics and
the dynamics dictated by the usual Dirac equation at different energy
scales. We address such a comparison as channel discrimination
problem and we quantify the difference between the two evolution with
the probability of error $p_e$ in the discrimination. We derive a
lower bound for $p_e$ as a function of momentum, mass and number of
the particles and the duration of the evolution.  Computing this bound
with experimentally achievable values, we see that automaton
evolution is undistinguishable from the one given by the Dirac field
equation. This result proves that, in the limit of input states with
vanishing momentum, the QCA evolution recovers the Dirac equation.
We notice that our analysis agrees with the works  \cite{PhysRevA.73.054302, strauch2007relativistic, PhysRevA.75.022322,
  PhysRevA.81.062340, di2012discrete} that studied the continuum limit,
i.e. when the lattice spacings and the time steps are sent to $0$, of
QWs in comparison with the Dirac or the Klein-Gordon equations
\footnote{We would like also to point out Ref.~\cite{arrighi2014dirac},
  which appeared after the first version of the present
  paper, where the authors proved convergence of the solution of the QW to the
  solution of the Cauchy problem for the Dirac equation}.

In order to gain insight about
the kinematics described by the QCA model we focus on one-particle
states that are smooth and have limited band in momentum.  Their
evolution can be approximated by a dispersive (momentum-dependent)
differential equation whose features can be easily compared with the
analogous expressions for the non-relativistic and relativistic cases.
By using this tool we will then study an elementary discrimination
experiment between the Dirac automaton evolution and the usual Dirac
one based on particle fly-time.

The line of research suggested by this paper explores the possibility that
quantum information processing underlies all of physics and is based on
the principle, for the first time proposed by Feynman
\cite{feynman1982simulating} and then refined by Deutsch \cite{deutsch1985quantum}, that every finite experimental
protocol is perfectly simulated by a finite quantum algorithm.
 It is immediate to see that the principle implies both
that the density of information is finite, and that the interactions
are local. 
The discreteness of the automaton framework
could also represent a possible way out of the typical problems affecting QFT originating
from the continuous background that still lack a satisfactory
interpretation (see
\cite{cao1993conceptual,auyang1995quantum,teller1997interpretive,
  cao2004conceptual}).  For example, in a QCA model there cannot be 
ultraviolet divergences since the presence of a discrete lattice 
implies a cutoff in momentum. 
 The QCA has an exact notion of observables, accommodates
localized states
\footnote{Since the automaton evolution is strictly causal any local
  excitation remains local during the evolution.}
 and measurements\footnote{The relevance of presenting
  QFT as a probabilistic theory about local measurements has been also
  the main focus of the so called {\em algebraic quantum field theory}
  \cite{haag1964algebraic,streater1964pct}.}, and is endowed with well defined
probabilistic interpretation and could lead to a coherent
unification of a hypothetical discrete Planck scale with the typical
Fermi scale of high-energy physics experiments. 
Finally, the field automaton 
is a physical model which is quantum {\em ab initio},  and is not derived by applying a
quantization procedure to classical field theory.
 
It is worth emphasizing that the difference between the QCA approach
and the discrete approach of lattice gauge theories is twofold.  On
the ``foundational'' side, our aim is 
to explore the idea whether it is possible to ground QFT on
a more fundamental QCA theory, and then recover the usual quantum fields as a large scale
approximation. Within this perspective, Lorentz covariance is supposed
to hold only in the
limit of small wave-vectors, whereas generally it is deformed
\cite{amelino2002relativity,amelino2001planck,PhysRevLett.88.190403,magueijo2003generalized}
while approaching the Planck scale\footnote{In the explorative approach of
this work we will describe the automaton dynamics in a fixed reference
frame while the study of boosted automata and the features of the
emerging spacetime have been the subject of another publication
\cite{bibeau2013doubly}}. On a more ``technical'' side the evolution of the
automaton is not given by a finite difference Hamiltonian or
Lagrangian as in lattice gauge theory. The quantum automaton is based
on a discrete and exactly causal unitary evolution and the Hamiltonian
has no longer any physical relevance. The same fact that there is no
Hamiltonian is the reason why the Fermion-doubling
\cite{nielsen1981no} is no longer an issue in the QCA framework (see
e.g. \cite{bialynicki1994weyl}).

We conclude this introductory section with a short outline of the paper.
In Sect.~\ref{s:Dirac}, after reviewing some generalities of QCAs and
QWs, we discuss the covariance of a QCA with respect to the
symmetry of the causal network and we derive a one dimensional linear
QCA from the assumptions of minimal internal
dimension, homogeneity, parity and time reversal invariance.
In Sect.~\ref{s:DiracB} we show how this automaton recovers the Dirac dynamics
for small masses and momenta.
Here, we set the problem by considering the probability of
error $p_e$ in the discrimination between the unitary channel corresponding
to the automaton evolution and the one which corresponds to the
evolution dictated by the Dirac equation. 
We obtain a lower bound for  $p_e$  in terms of the
mass of the field, the number and the momentum of the particles,
and the duration of the evolution.
Then in Sect.~\ref{s:analytic} we present an analytical approximation method for evaluating the
automaton evolution for one-particle states which are smooth in momentum and
with limited bandwidth. Then we derive a dispersive (momentum-dependent) differential
equation, which approximate the QCA evolution. We compare computer simulations with
the analytic approximation, and provide the leading order corrections
to the Dirac equation.
After discussing possible ways of testing of the theory, like the
effects of the automaton evolution on wave-packets fly-times, we conclude the
paper with future perspectives.

\section{The one-dimensional Dirac automaton}\label{s:Dirac}

\subsection{One dimensional field QCA and quantum walks}

In this section, we present some
generalities about QCA in one dimension and we review the notion
of linear QCA and its connection with the one of quantum walk.

A one-dimensional QCA describes the discrete time unitary local
evolution of quantum systems on the one-dimensional lattice $\Z$.
Since we want to apply this model of evolution to quantum
fields, any site $x \in \mathbb{Z}$ will correspond to a
Bosonic or Fermionic quantum field operator $\psi(x)$ located at the
same position.  If the field has $\Lambda $ internal degrees of
freedom the operators
\begin{align}
\{\psi_a(x)\}_{ a\in\mathrm{A}},\qquad \mathrm{A}=\{1,\ldots,\Lambda\}
\end{align}
will denote the generators of the field local algebra $\mathcal{F}_x$ that
satisfies the usual commutation, respectively anticommutation, rules
$
[\psi_a(x) , \psi_b(x)]_\pm=[\psi_a^{\dag}(x) , \psi_b^{\dag}(y)]_\pm=0$, $[\psi_a^{\dag}(x),\psi_b(y)]_\pm=
  \delta_{xy}\delta_{ab}$.
  The automaton corresponding to the one-step update of the field is
required to preserve the above relation. In the usual QFT both the Fermionic and the Bosonic algebra's
structure is preserved with the field evolving by a unitary operator $U$
\begin{align}\label{eq:field-evolution}
  \psi(x,t+1)=U^\dag\psi(x,t)U.
\end{align}
If we restrict to the evolution of free, i.e. non interacting, fields
the evolution in Eq. \eqref{eq:field-evolution}
is linear in the field, namely we have
\begin{align}\label{eq:field-evolution2}
  \psi_a(x,t+1)=\sum_{y\in\Z,\,b\in\mathrm{A}}U^{ab}_{xy}\,\psi_b(y,t)
\end{align}
for some complex coefficients $U^{ab}_{xy}$. 
Upon introducing the vector field $\v{\psi}$
\begin{align}
  \label{eq:1}
\v{\psi} := 
  (
    \dots, 
    \v\psi(x),
    \v\psi(x+1),
    \dots
  )^\top,
  \quad \v{\psi} (x) :=
  (
\psi_1(x),
\dots,
\psi_{\Lambda}(x)
)^\top,
\end{align}
where each $\v\psi(x)$ is also a vector with $\Lambda$ components
corresponding to the internal degrees of freedom of the field, we have
the equality $\v{\psi}(t+1) = \v{U} \v{\psi}(t)$ where $\v{U}$ is the unitary matrix
$\v{U}\v{U}^\dag=\v{U}^\dag \v{U}=I$ having entries $U^{ab}_{xy}$ according to
Eq.~\eqref{eq:field-evolution2}. 

If we want the evolution of Eq. \eqref{eq:1} to be local,
$\v{\psi}(x,t+1)$ must be a linear combination of the field on few
neighboring sites $\Neigh_x\subset\Z$ at time step $t$, that is
\begin{align}\label{eq:band}
  \psi_a(x,t+1) =\sum_{y\in\Z,\, b\in\mathrm{A}}
  U^{ab}_{xy}\,\psi_b(y,t),\qquad
  U^{ab}_{xy}=0\quad \forall y\notin\Neigh_x.
\end{align}
The map $\v{U}$ represents then a linear QCA with ``cell structure''
$\mathcal{F}_x$ and \emph{neighborhood scheme} $\Neigh_x$.  In the
following we consider automata with {\em nearest neighborhood scheme},
namely $\Neigh_x=\{x-1,x,x+1\}$ (see left Fig.~\ref{fig:automaton}) (the
next-neighboring interaction is not an assumption by itself, since it
is always possible to reduce to such a case by grouping a periodic
pattern of the network into a single node of the automaton) and
satisfying \emph{translational invariance}, say $\v{U}$ must commute with
the shift operator
  \begin{align} [\v{U},S_1]=0,\quad S_1: \v{\psi}(x)\rightarrow\v{\psi}(x+1).
\end{align}
This implies that the only non zero entries of the matrix $\v{U}$ are
$\v{U}_{y,y\pm 1}=\v{U}_{\pm1}$ and $\v{U}_{y,y}=\v{U}_{0}$ and that 
$\v{U}$ has the
simple band diagonal form
\begin{align}
\label{eq:translation-invariance-matrix}
\v{U}=\sum_{x\in\{\minus 1,0,1\}} \v{U}_{x}\otimes S_{x}=\begin{pmatrix}
\ddots&\ddots&\ddots&{}&{}&{}&{}\\
{}&\v{U}_{\minus 1}&\v{U}_{0}&\v{U}_{1}&{}&{}&{}\\
{}&{}&\v{U}_{\minus 1}&\v{U}_{0}&\v{U}_{1}&{}&{}\\
{}&{}&{}&\v{U}_{\minus 1}&\v{U}_{0}&\v{U}_{1}&{}\\
{}&{}&{}&{}&\ddots&\ddots&\ddots\\
\end{pmatrix}
\end{align} 
where the $\v{U}_x$'s $\Lambda\times\Lambda$ are called
\emph{transition matrices}, while $S_1,S_{\minus 1}, S_0$ correspond
respectively to the right shift, left shift and the identity. 

This framework is formally equivalent to a quantum walk 
on the
Hilbert space $\mathbb{C}^{\Lambda}\otimes l_2(\Z)$ with
$\mathbb{C}^{\Lambda}$ the particle internal Hilbert space.
A quantum walk is the generalization in the quantum framework
of the common notion of random walk and it was introduced for the
first time in Ref. \cite{aharonov1993quantum} (for a review on QWs see
e.g. Ref \cite{reitzner2011quantum} and references therein).
It is
known that a QCA restricted to the one-particle sector corresponds to
a QW. However, when the evolution is linear as in the present case,
the one-particle dynamics fully specifies the
evolution of many particles.
By reversing the line of reasoning one  can realize that 
a linear field-QCA is obtained by ``promoting''
the state $\ket{\v{\psi}}$ in the
usual QW framework
to a vector of
field operators $\v{\psi}$ (see the ``second quantization'' of a QW of
Ref.~\cite{schumacher2004reversible}).

For convenience, as usual in the literature
\cite{ambainis2001one,knight2004propagating,valcarcel2010tailoring,ahlbrecht2011asymptotic,reitzner2011quantum},
we will study the dynamics of the field automaton in the momentum
representation.  For the field operator we have
\begin{align}\label{eq:fourier-convention}
\v{\psi}(k):=\frac{1}{\sqrt{2\pi}}\sum_{x\in\Z} e^{-ikx}\v{\psi}(x),\quad   k\in[-\pi,\pi],
\end{align}
where with little abuse of notation we utilize the variable name $k$
to denote the Fourier transform of any function of $x$. Notice that
the automaton model is naturally band-limited $k\in[-\pi,\pi]$ and
periodic in momenta due to the discreteness of the lattice.
The automaton in the momentum space is then given by
\begin{align}\label{eq:automaton-Uk-0}
 \v{U}= \int_{-\pi}^{\pi} \d{k} \v{U}(k)\otimes\ketbra{k}{k},\qquad \v{U}(k)=\sum_{x\in\{-1,0,1\}} \v{U}_{x}e^{-ikx},
\end{align}
and we can define the Hamiltonian $\v{H}$ that describes the
automaton evolution for continuous times, interpolating between
time-steps, namely
\begin{align}\label{eq:Hunp-0}
\v{H}=\int_{-\pi}^\pi \d{k} \v{H}(k)\otimes\ketbra{k}{k},\qquad \v{U}^t=
\exp(-i\v{H}t).
\end{align}
Upon diagonalizing the unitary $\v{U}(k)$ we get the automaton
one-particle eigenvalues and eigenvectors
\begin{align}
u_k(s)=e^{-is\omega(k)},\qquad \ket{s}_k,\qquad s=\pm,
\end{align}
with $\omega(k)$ the automaton dispersion relation.

The field automaton $\v{U}$ generally operates on the vector field
$\v\psi$ which describes an arbitrary number of particles.  The vacuum
state for the automaton is defined as the state $\ket{\Omega}$ such
that
\begin{align}
 \psi_s(k)\ket{\Omega}=0\qquad\forall s=\pm,\quad\forall k\in[-\pi,\pi].
\end{align}
Up to now, we have not specified the nature Fermionic/Bosonic of the field. Here
we will focus only on the Fermionic case of anticommuting field. A $N$-particle state can be
obtained by acting with the field operator on the vacuum as follows
\begin{equation}
|N,\v{k},\v{s}\>=\left(\prod_{i=1}^N\psi_{s_i}^\dag(k_i)\right)\ket{\Omega}.
\end{equation}
Specifically, for $N=1$ particle eigenstates of $\v{U}$, we write
\begin{equation}
\psi^\dag_s(k)\ket{\Omega}=\ket{s}_k|k\>,
\end{equation}
whereas for $N=2$ we have
$
\psi^\dag_{s_1}(k_1)\psi^\dag_{s_2}(k_2)\ket{\Omega}=\ket{s_1}_{k_1}\ket{s_1}_{k_2}|k_1,k_2\>
$
where $|k_1,k_2\>=-|k_2,k_1\>$, and so forth for $N>2$. The corresponding eigenvalues of the
(logarithm of ) $\v{U}$ are
$\omega(N,\v{k},\v{s})=\sum_{i=1}^{N}s_i\,\omega(k_i,m)$.

\subsection{Derivation of the one-dimensional Dirac automaton}

In this section we present the derivation of
simplest field automaton, here denoted Dirac QCA, that is covariant
with respect to the symmetries of the one-dimensional causal network
and that exhibits a non trivial evolution.

\begin{figure}[t!]
\centering
\includegraphics[width=0.38\textwidth]{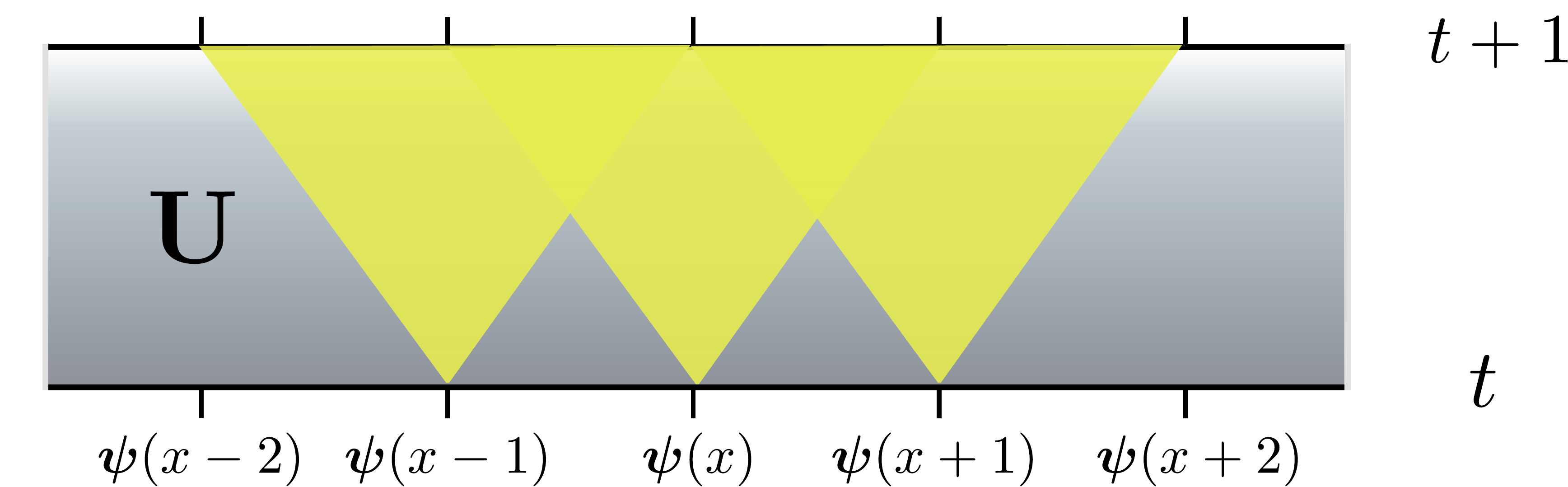}\qquad
\includegraphics[width=0.38\textwidth]{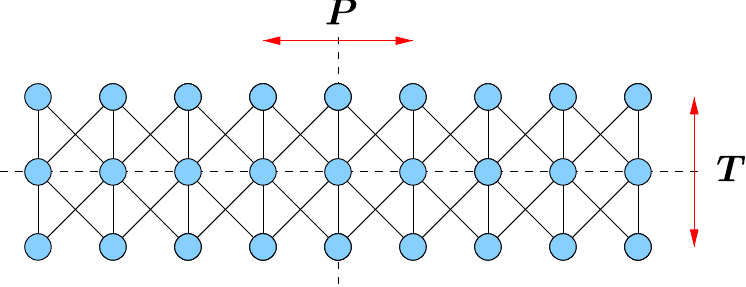}
\caption{{\bf Left figure:}Illustration of a one-dimensional QCA unitary step. Each site of the
  lattice $x$ corresponds to a quantum field evaluation $\v{\psi}(x)$. The field operator at site
  $x$ interacts with the field $\v{\psi}(x\pm 1)$ at neighboring sites.  In the case of the Dirac
  automaton the field operator has two components (see text).
{\bf Right figure:} Schematic of the three time steps causal network corresponding to a one-dimensional quantum
  cellular automaton with next-neighboring interaction. The topology of the network is left
  invariant by the mappings $x\mapsto -x$ and $t\mapsto -t$ and the dynamics of the automaton is
  assumed to be parity ($\v{P}$) and  time reversal ($\v{T}$) invariant (see Eqs.  \eqref{eq:parity}
  and \eqref{eq:time-reversal}). }\label{fig:automaton}
\end{figure}

In general, the lattice of an automaton is endowed with certain
discrete symmetries. The one-dimensional lattice $\Z$ only exhibits
parity symmetry (see right Fig.~\ref{fig:automaton}), corresponding to the lattice reflection with respect
to some site (and the time reversal symmetry if we consider the causal
network given by the automaton evolution)  . The invariance of the automaton dynamics with respect to a discrete 
symmetry of the lattice, given by a group G  whose elements
are linear functions $g:\Z\rightarrow\Z$, is satisfied if the QCA is \emph{covariant} under a unitary representation $\{
\v{O}_g\}$ of $G$ on the field local algebra $\v{O}_g:\mathcal{F}_x\rightarrow\mathcal{F}_{g(x)}$, namely
\begin{equation}\label{eq:covariance}
\v{U}=\sum_{x\in\{\minus 1,0,1\}} \v{O}_g\v{U}_x \v{O}^\dag_g\otimes
S_{g(x)},\qquad \forall g\in G.
\end{equation}

In \ref{a:derivation} we derive the simplest covariant
one-dimensional automaton that exhibits a non-trivial (non-identical)
evolution.  We can summarize our assumptions as follows:
\begin{enumerate}[(i)]
\setlength{\itemsep}{1pt}
\setlength{\parskip}{0pt}
\setlength{\parsep}{0pt}
\item\label{1} Unitarity of the evolution;
\item\label{2} Translation invariance;
\item\label{xx} Covariance under parity $x\mapsto -x$; 
\item\label{tt} Covariance under  time-reversal $t\mapsto -t$;
\item\label{minimal} Minimal internal dimension $\Lambda$ for a
  non-identical evolution.
\end{enumerate}

The first two assumptions are already contained in the definition
itself of translational invariant QCA which has the general form given
in Eq.~\eqref{eq:translation-invariance-matrix}. From the band
diagonal form of the unitary $\v{U}$ it is immediate to see that the
unitarity condition $\v{U}\v{U}^\dag=\v{U}^\dag \v{U}=I$ is equivalent
to the following constraints on the transition matrices
\begin{equation}
  \label{eq:unitarity}
\begin{split}
\v{U}_1\v{U}_1^\dag+\v{U}_{\minus 1}\v{U}_{\minus 1}^\dag+\v{U}_0\v{U}_0^\dag=I\,\qquad
 \v{U}_0\v{U}_1^\dag+\v{U}_{\minus 1}\v{U}_0^\dag=0\,,\qquad \v{U}_{\minus 1}\v{U}_1^\dag=0\;.
\end{split}
\end{equation}
Assumptions \eqref{xx} and \eqref{tt} require the automaton to preserve the
symmetries of the one-dimensional causal network (in right
Fig.~\ref{fig:automaton}). The covariance for parity symmetry
 can be expressed as in Eq.~\eqref{eq:covariance} where the 
parity transformation $g(x)=-x$ has to be represented on the field local algebra via a unitary matrix $\v{P}$ such that
\begin{align}\label{eq:parity}
\v{U}=\v{P}\v{U}_1\v{P}^\dag\otimes S_{\minus 1}+\v{P}\v{U}_{\minus 1}\v{P}^\dag\otimes S_1+\v{P}\v{U}_0\v{P}^\dag
\otimes I.
\end{align}
Similarly, we impose the covariance for time reversal, which is not a symmetry of the lattice but of the full causal 
network, asking that  
\begin{align}\label{eq:time-reversal}
\v{U}=(\v{T}\otimes I)\v{U}^\dag(\v{T}^\dag\otimes I)=
\v{T}\v{U}_1^\dag \v{T}^\dag\otimes S_{\minus 1}+\v{T}\v{U}_{\minus 1}^\dag\v{T}^\dag\otimes S_1+\v{T}\v{U}_0^\dag\v{T}^\dag\otimes I,
\end{align}
for some anti-unitary operator $\v{T}$ (see \ref{a:derivation} for
the anti-unitarity of time reversal).

For $\Lambda=1$ the only translational invariant QCA satisfying parity
invariance is the identical one $\v{U}=I$ (see 
\ref{a:derivation}) as already proved by Meyer in the context of
quantum lattice gases \cite{meyer1996quantum}. Next, we have the case
$\Lambda=2$. In this case we find (see \ref{a:derivation})
that all the QCAs satisfying the conditions above and are then unitarily
equivalent to the following automaton
\begin{align}\label{eq:U}
\v{U}=\begin{pmatrix}
  n S_{-1} &-im\\
  -im & n S_{1}
\end{pmatrix},\quad n,m\in\Reals^+,\quad n^2+m^2=1,\,\qquad \v{\psi}(x) :=
  \begin{pmatrix}
\psi_L(x) \\
\psi_R(x) \\
\end{pmatrix},
\end{align}
where we named the two components of the field $\psi_R$ and $\psi_L$
{\em left} and {\em right} modes. Among the class of unitary
equivalent QCAs, we have chosen the one whose expression reproduces
the Dirac equation in the Weyl representation as a finite difference
equation.  The unitarity constraint $ n^2+m^2=1$ in Eq.~\eqref{eq:U}
forces the parameter $m$ to be $m\in[0,1]$ \footnote{We will see in
  the next Section that in a certain limit the Dirac automaton
  evolution mimics the solutions of the Dirac evolution and the
  parameter $m$ will play the role of a the mass.}.  

In the momentum space (see Eq.~\eqref{eq:automaton-Uk-0}) the Dirac automaton is given by
\begin{align}\label{eq:automaton-Uk}
\v{U}(k)=\sum_{x\in\{-1,0,1\}} \v{U}_{x}e^{-ikx}=
 \begin{pmatrix}
    n e^{ik} & -i m\\
-i m & n e^{-ik}
\end{pmatrix},\qquad
\end{align}
and, upon its diagonalization, it is easy to derive the Hamiltonian of 
Eq.~\eqref{eq:Hunp-0}
\begin{align}\label{eq:Hunp}
\v{H}(k)=\frac {\omega}{\sin(\omega)} \begin{pmatrix}
    -n\sin(k)&m\\
    m& n\sin(k)
\end{pmatrix},
\end{align}
with $\omega(k,m)$ the automaton dispersion relation
\begin{equation}\label{eq:disp}
\omega(k,m)=\arccos(\sqrt{1-m^2}\cos(k)).
\end{equation}
In 1d we have that $\omega(k,m)$ is an increasing function of $|k|$,
and then there is no Fermion doubling, namely no state other than for
$k=0$ corresponding to a minimum of the energy $\omega(k,m)$ (for
dimension greater than one, the dispersion relation can be as well
made monotonic continuous by exploiting the multi-valued nature of the
dispersion relation \cite{PhysRevA.90.062106}, as pointed out in Ref.~\cite{bialynicki1994weyl}). 
The eigenvalues and the eigenvectors of the unitary matrix $\v{U}(k)$
in Eq.~\eqref{eq:automaton-Uk} are given by
\begin{align}\label{eq:eigenstates}
u_k(s)=e^{-is\omega},\;\;\ket{s}_k:=\tfrac{1}{\sqrt{2}}
\begin{bmatrix}
\sqrt{1-sv}\\s\sqrt{1+sv}
\end{bmatrix},\quad s=\pm,
\end{align}
in terms of the automaton dispersion relation (\ref{eq:disp}) and the group velocity
$v=\partial_k\omega$.

As we will see in Section \ref{s:analytic}, in analogy with the Dirac theory, the eigenvalues with $s=1$ in
Eq.~\eqref{eq:eigenstates} correspond to positive-energy particle
states, whereas the eigenvalues with  $s=-1$ correspond to
negative-energy anti-particle states.  The most general state
$\ket{\psi}$ is thus a superposition of a positive and a negative
energy state, i.e.  $\ket{\psi_+} + \ket{\psi_-}$, and typical aspects
of the Dirac-field dynamics, such as the Zitterbewegung and the Klein
paradox, are also dynamical feature of the Dirac automaton as shown by
the authors in a more recent paper \cite{bisio2013dirac}.

\begin{figure}
\centering
\includegraphics[width=.53\textwidth]{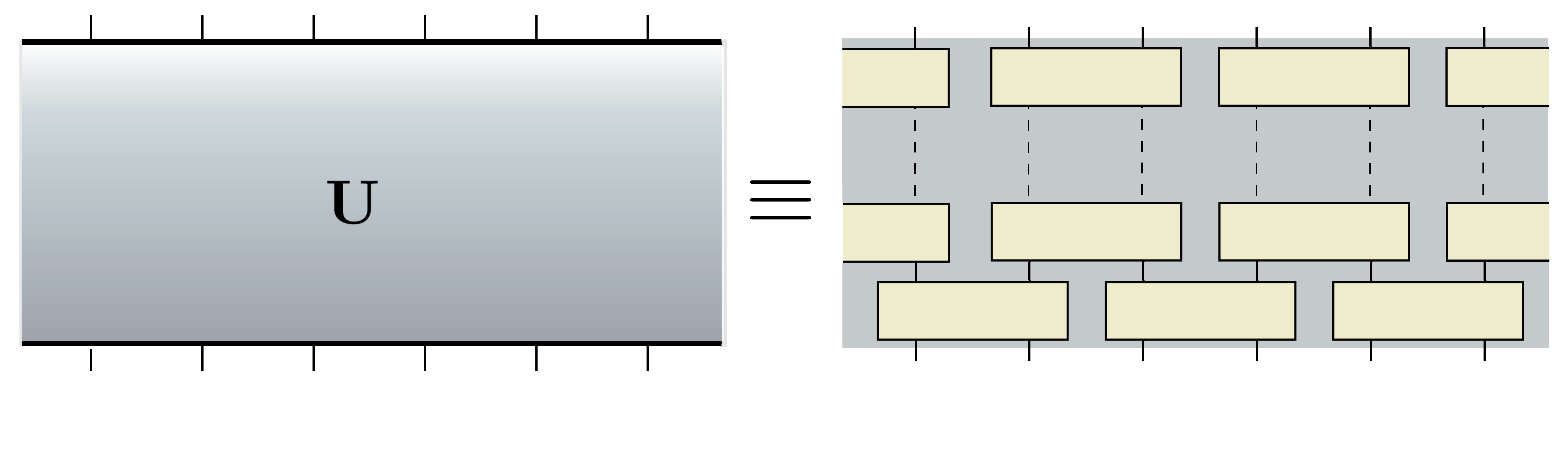}\qquad
\includegraphics[width=.38\textwidth]{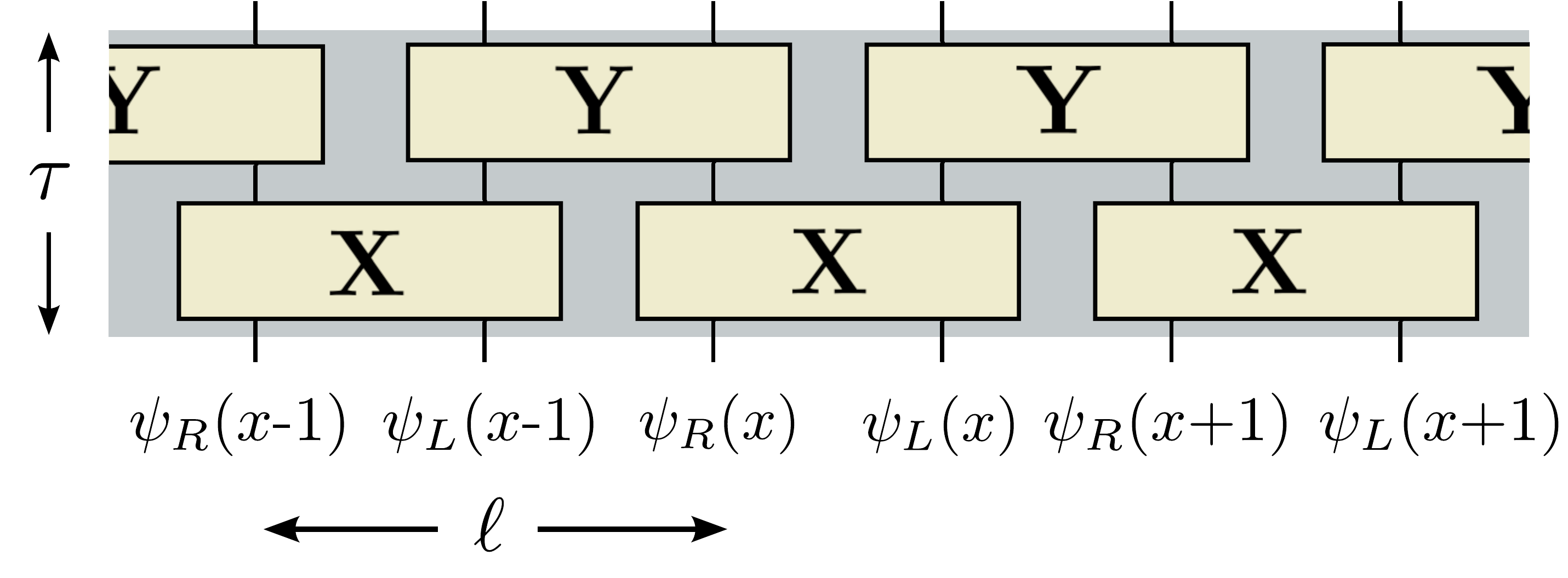}
\caption{{\bf Left figure:} local implementation of the generic one
  dimensional automaton $\v{U}$.  {\bf Right figure:} the local
  implementation of the Dirac automaton \eqref{eq:U}
  with $\v{X}=-i\sigma_1$,
  $\v{Y}=mI+in\sigma_1.$.}\label{fig:localizability}
\end{figure}

Notice that in the derivation of the automaton our assumptions imply a
minimal internal dimension $\Lambda=2$ for a non-identical
evolution. This means that it is not possible to consider an automaton
having just an internal degree of freedom--say a scalar field.
Moreover, although it is not the focus of this work, it is interesting
to notice that as a byproduct of the assumptions leading to the Dirac
automaton we also have its localizability, namely the possibility of
decomposing the unitary $\v{U}$ in a number of more elementary gates
involving only neighboring systems as shown in the left
Fig.~\ref{fig:localizability}. This is the so-called Margolus scheme
\cite{toffoli1987cellular}. It is well known from the cellular
automata and walks theory that the locality of the automaton does not
ensure the existence of a local implementation (typical examples of
local but non localizable automata are the right and left shifts, which
do not satisfy parity). Werner \emph{et al.}  proved
\cite{gross2012index} that a necessary and sufficient condition for
the localizability of a QW is that $\det{(\v{U}(k))}=$ const, where
$\v{U}(k)$ is the momentum representation of the walk.  As already
noticed the one-particle sector of the Dirac field automaton $\v{U}$
coincides with a walk (see
Eq.~\eqref{eq:translation-invariance-matrix}) and we can exploit the
above result. In the Dirac case it is $\det{(\v{U}(k))}=n^2+m^2=1$
(see Eq.~\eqref{eq:automaton-Uk}) which shows the localizability of
corresponding unitary evolution. Moreover, as shown by Arrighi
\emph{et al.}  \cite{arrighi2011unitarity}, a localizable $d$
dimensional QCA can be locally implemented using $2^{d}$ layers of
quantum gates and then by just two layers in the 1d case.  For the
one-dimensional Dirac automaton \eqref{eq:U} we have the local
implementation shown in the right Fig.~\ref{fig:localizability} and
the local gate $\v{X}$ and $\v{Y}$ are as follows \cite{darianopla}
\begin{align}
\v{X}=-i\sigma_1, \qquad
\v{Y}=m I+in\sigma_1.
\end{align}

\section{Recovering the Dirac dynamics}\label{s:DiracB}
By interpreting the parameters $k$ and $m$ of 
the Dirac automaton as momentum and mass,
it is reasonable to expect that the usual
kinematics of the Dirac equation
\begin{align}\label{eq:dirac-hamiltonian}
  i\partial_t \v{\psi}(k,t) = \v{H}_{\rm D}(k)
\v{\psi}(k,t)
\qquad \mbox{where}\quad
 \v{H}_{\rm D}(k)=\begin{pmatrix}
    -k&m\\
    m&k
\end{pmatrix}.
\end{align}
is recovered in the small momenta ($k \to 0$) and small mass ($m \to
0$) regime.  More precisely, one would say that it is not possible to
tell the difference between the automaton evolution $\ket{\psi(t)} =
U_A^{t} \ket{\psi(0)}$ and the evolution given by Dirac Hamiltonian
$\ket{\psi(t)} = U_D^{t} \ket{\psi(0)}$ ($U_D^{t}$ is the unitary
evolution given by the Dirac Hamiltonian) as far as the mass $m$ is small
and the momentum of initial state $\ket{\psi}$ is small. This idea can
be rigorously recast in terms of a discrimination problem between two
black boxes. The scenario can be described as follows.  An
experimentalist is given a black box that can be either the automaton
(box $\rm A$) or the usual Dirac equation (box $\rm D$) with equal
probability, and he is asked to guess which box.  The most general
experiment which discriminates between two unitary evolutions amounts
to the following three steps procedure\footnote{In general, the
  optimal discrimination needs entangled states, but for the case
  of two unitary evolutions this is not necessary \cite{d2001using}.}: i) preparing a
quantum state $\rho$, ii) apply the unknown unitary evolution $U_X$
($X = A,D$) iii) perform a two outcome measurement on the output
state: the two outcomes $A$ and $D$ correspond to the two possible
evolutions.  The measurement is described by a positive operator
valued measure (POVM) $\v{P} = \{ P_A, P_D\}$, where $P_A$ and $P_D$
are positive operators on $\mathcal{H} \otimes \mathcal{K}$ which
satisfy $P_A + P_D = I$, $I$ denoting the identity.
Then the probability of error reads
\begin{align}
  \label{eq:proberr2}
  p_{e}(P_A, \rho) =\frac12 \Tr[P_A(( U_D \otimes I)\rho( {U_D}^\dagger \otimes I) - (U_A \otimes I) \rho( {U_A}^\dagger \otimes I))].
\end{align}
It is clear from this scenario that the minimum of the probability of
error over all the possible experiments is a well defined measure of
how much the models $\rm A$ and $\rm D$ are far apart.  Minimizing expression
\eqref{eq:proberr2} over all the possible experiments entails a
minimization over the set of the POVM's and the set of the available
states.  The minimization over the POVM set gives
\cite{helstrom1976quantum}:
\begin{align}
  \label{eq:6}
  \inf_{0 \leq P_A \leq I} p_{e} =\frac12 -\frac12 || U_D \rho
  {U_D}^\dagger - U_A \rho {U_A}^\dagger||_1
\end{align}
where $|| \sigma ||_1$ denotes the trace norm $|| \sigma ||_1 =
\Tr[\sqrt{\sigma^\dagger \sigma}]$.
If we now  set bounds on the number of
particles $N\leq\bar{N}$ and their momentum $k\leq\bar{k}$,
the minimization over the admissible input states $\rho$ is:
\begin{equation}\label{eq:supoverstate}
  \bar{p}_e= \tfrac{1}{2} -\tfrac{1}{2}
  \sup_{\rho\in{\mathcal {T}_{\bar{k},\bar{N}}}}|\!|U_{\rm A}\rho U_{\rm A}^\dag
    -U_{\rm D}\rho U_{\rm D}^\dag|\!|_1,  
\end{equation}
where $\mathcal {T}_{\bar{k},\bar{N}}$ denotes the set
\begin{equation}\label{eq:termostate}
\rho\in \mathcal {T}_{\bar{k},\bar{N}}\text{ iff }
\Tr[\rho N_{\bar{k}}] =\Tr[\rho P_{\bar{N}}]=0
\end{equation}
where $P_{\bar{N}}$ is the projector on the $N>\bar{N}$-particles
sector and $N_{\bar{k}}$ is the operator that counts the number of
particles with momentum $|k|>\bar{k}$, i.e $N_{\bar{k}}=\int_{|k|>
  \bar{k}} \;\d{k}  \v{\psi}^{\dagger}(k)\v{\psi}(k)$.  

In
\ref{a:large-scale} we evaluate a lower bound for $ \bar{p}_e$
probability of error, which is given by
\begin{align}
  \label{eq:ilboundperpe}
  \bar{p}_e\geq \tfrac12 - \tfrac12
  \sqrt{1-\cos^2(g(\bar{k},m,\bar{N},t))}
\end{align}
where
\begin{equation}
\begin{split}
  & g(\bar{k},m,\bar{N},t):= \bar{N}\arccos\left(
    \cos({\bar{\alpha}}t) -{\bar{\beta}} \right)\label{eq:gfunction}\\
  &\bar{\alpha}:= \max_{k \in\{0,\bar{k}\}} |\omega_{\rm D} - \omega|, \qquad
  \bar{\beta}:= \max_{k \in\{0,\bar{k}\}} \left|\frac12 \left( 1 - v v_{\rm D} -  \sqrt{(1-v^2)(1-v_{\rm D}^2)} \right)\right|
\end{split}
\end{equation}
and $v$ (see Eq.~\eqref{eq:drift}) and $v_{\rm D}=k/\sqrt{k^2+m^2}$ the automaton and the Dirac
drift coefficients. The bound in Eq. \eqref{eq:ilboundperpe}
 explicitly quantifies the similarity between the evolution described
by the automaton of
Eq. \eqref{eq:U} and the evolution described by the Dirac equation.
Moreover, this result is exact (i.e.  it does not depend on any
approximations),
easily computable (it is an explicit function of $m,\bar{k},\bar{N},t$),
and provide an experimentally meaningful numerical value (since $p_e$ is the probability of an
experiment). 

A simplified version of the bound in the $k,m\ll 1$ regime can be
obtained by expanding in series the function $g$ in Eq.~\eqref{eq:gfunction} near $m = \bar{k} =
0$. Truncating the expansion at the leading order and neglecting a small constant term we have
\begin{align}
  \label{eq:5}
  g(m,\bar{k},\bar{N},t) \approx \frac{1}{6} m^2 \bar{k} \,
  \bar{N} \, t.
\end{align}
By putting $\bar{p}_e=0$, corresponding to $g(m,\bar{k},\bar{N},t)=\pi/2$, we obtain the minimum
time required for discriminating perfectly between the automaton and the Dirac evolution
\begin{align}\label{eq:time-scale}
  t_{min}(m,\bar{k},\bar{N})\approx 3\pi\frac{1}{m^2\bar{k}\bar{N}}.
\end{align}
Notice that this is an in-principle result, without any specification of the actual
apparatus needed to achieve it. For a proton with $\bar{k}=k_{CR}\approx
10^{-8}$ (as for order of
magnitude, we consider numerical values corresponding to ultra high
energy cosmic rays (UHECR) \cite{takeda1998extension}) we have
\begin{align}
  t_{min}(m_p,k_{CR},1)\approx 3\pi 10^{46}\:\text{Planck
    times}\:\approx 10^3s,
\end{align}

\begin{figure}[t]
\begin{minipage}[b]{7cm}
\centering 
\includegraphics[width=6cm]{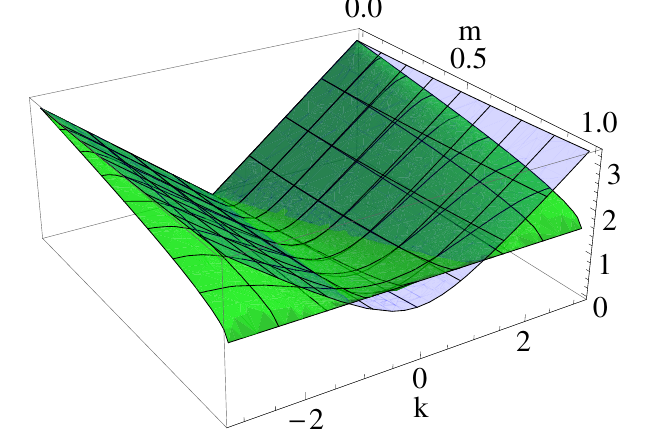}
\end{minipage}
\begin{minipage}[b]{8cm}
\centering
\includegraphics[width=3.5cm]{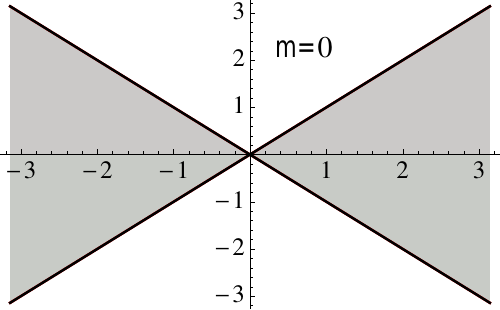} 
\includegraphics[width=3.5cm]{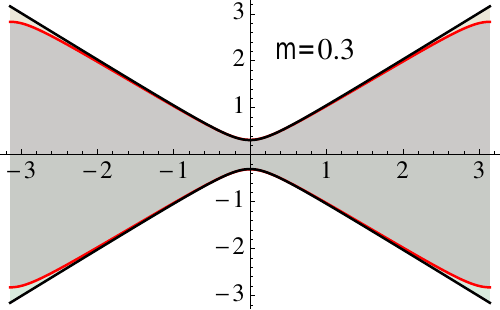}\\
\includegraphics[width=3.5cm]{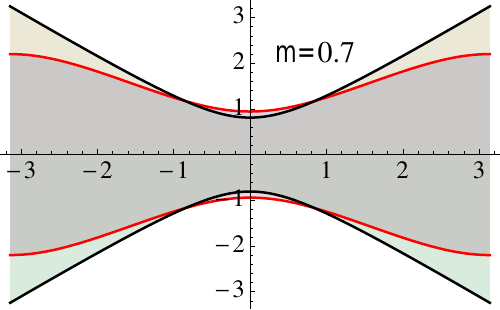}
\includegraphics[width=3.5cm]{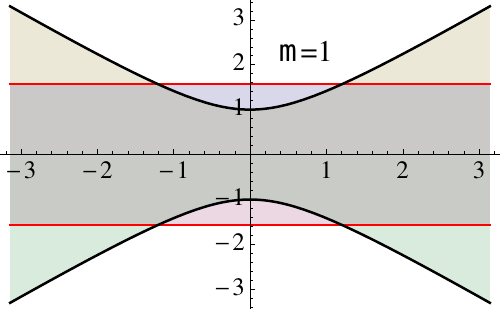}
\end{minipage}
\caption{(Colors online) Comparison between the dispersion relations $\omega(k,m)$ of the Dirac
  automaton and of the Dirac equation, in Eqs. \eqref{eq:disp} and \eqref{eq:power-expansion2},
  respectively. In the top figure the dispersion relation is plotted
  versus the adimensional mass
  $m\in[0,1]$ and momentum $k\in[-\pi,\pi]$ ($m=1$ corresponds to the Planck mass). The green
  surface represents the automaton, whereas the blue the Dirac one. In the bottom figures
  $\omega(k,m)$ is plotted versus $k$ for four values of $m$ (the red line corresponds to the
  automaton, whereas the black one is the Dirac's). We can see that the two dispersion relations
  coincide for small masses and momenta, and the larger the mass the smaller the overlap region
  around $k=0$.}\label{fig:3D-disp-rel}
\end{figure}

The consistency of our result can be checked by power expanding
the Hamiltonian of Eq.~\eqref{eq:Hunp} and the dispersion relation of Eq.~\eqref{eq:disp} in the limit
of $k,m \to 0$, 
\begin{align}\label{eq:power-expansion2}
  \begin{split}
  \v{H}_{\rm A}(k)& \simeq \v{H}_{\rm D}(k)+\frac{m}{3}
  \begin{pmatrix}
    mk&\tfrac{1}{2}(k^2+m^2)\\
    \tfrac{1}{2}(k^2+m^2)&-mk
  \end{pmatrix}
\\
\\
  \omega_{\rm A} & \simeq \omega_{\rm
    D}\left(1-\frac{m^2}{6}\frac{k^2-m^2}{k^2+m^2}\right)
\qquad 
 \omega_{\rm D} := \omega_{\rm D}(k,m) = \sqrt{k^2 + m^2}.
\end{split}
\end{align}
and see that the leading terms are the the Dirac Hamiltonian
and the usual relativistic dispersion relation
(see also Fig.~\ref{fig:3D-disp-rel}).

The result of this section supports our interpretation of the
parameters $k$ and $m$ of the automaton with the momentum and the mass
of the Dirac field, respectively. Since the the typical rest masses
and momenta of particle physics experiments are many order of
magnitude smaller than the Planck mass and the Planck momentum also
the approximations \eqref{eq:5} and \eqref{eq:power-expansion2} are
justified.

One can say that the bound \ref{eq:ilboundperpe}  extends
``outside the limit'' the results of
Refs.~\cite{bialynicki1994weyl,strauch2007relativistic,PhysRevA.73.054302}
which compared the quantum walks model with the Dirac equations. 
Here we also have the additional bonuses that the many particle case
is included and that  the bound is explicitly written in terms of physical quantities like momentum, mass
and number of the particle and is given in terms of an experimental
meaningful quantity, i.e. the probability of error in a quantum
channels discrimination procedure.

\section{The one particle-sector of the Dirac automaton}\label{s:analytic}

In this section we explore the behavior of one particle states of the
Dirac QCA. In particular we will consider initial
states whose momentum distribution is smoothly peaked around some
$k_0$, namely
\begin{equation}
\label{eq:smoothstate}
\ket{\v{\psi} (0) } = \int_{-\pi}^{\pi}\,\df{k}{\sqrt{2\pi}} 
g(k,0) \ket{s}_k \ket{k}, \qquad s=\pm,
\end{equation}
where $g(k,0)\in C^\infty_0[-\pi,\pi]$ is a smooth function satisfying the bound
\begin{equation}\label{eq:smoothstate2}
\frac{1}{2\pi}\int_{k_0-\sigma}^{k_0+\sigma} \d{k}
| g(k,0) |^2 \geq 1-\epsilon,
\quad \sigma , \epsilon >0,
\end{equation}
and the two-component vector $|s\>_k$ is defined in Eq.~\eqref{eq:eigenstates}.

At time $t$ and in the position representation, the state in Eq.~\eqref{eq:smoothstate} can be
written as
\begin{align}
&\ket{\psi(t)} = \sum_x \ket{\psi (x,t) } \ket{x}\qquad 
\ket{\psi (x,t) } := e^{i(k_0 x - s\, \omega_0 t) } \ket{{\phi}(x,t)} 
\nonumber\\
&\ket{{\phi}(x,t)} := \int_{-\pi}^{\pi}\,\df{k}{\sqrt{2\pi}} 
e^{i(Kx-s\Omega(k,m)t) }g(k,0) \ket{s}_k 
\label{eq:smoothstatetimet}
\end{align}
where we posed 
$K = k-k_0$ and $\Omega(k,m) = \omega(k,m)-\omega_0$, with $\omega_0 = \omega(k_0,m)$. 
It is convenient to take $x,t$ to be real-valued continuous variable by extending the Fourier transform
in Eq.~\eqref{eq:smoothstatetimet} to real $x,t$. We derive the integral in Eq.~\eqref{eq:smoothstatetimet}
with respect to $t$, and expand $\Omega$ vs $k$ around $k_0$ up to
the second order. Then, taking the resulting derivatives with respect to $x$ out of the integral
(using the dominated derivative theorem), we obtain the following
dispersive differential equation with drift
\begin{equation}
\label{eq:dscvs}
i\partial_t \ket{\tilde{\phi}(x,t)}= s\left(i v
  \frac{\partial}{\partial x} -\frac{1}{2}D \frac{\partial
    ^2}{\partial x^2} \right) \ket{\tilde{\phi}(x,t)},
\end{equation}
with the drift constant $v$ and the diffusion constant $D$ depending on $k$ and $m$ as follows
\begin{align}
\label{eq:drift}
v:=\sqrt{\frac{1-m^2}{1+m^2\cot^2(k_0)}},\qquad D:=\frac{\sqrt{1-m^2}m^2\cos{(k_0)}}{(\sin^2(k_0)+m^2
  \cos^2(k_0))^{\frac32}},
\end{align}
and with the identification of the initial condition
$\ket{\tilde{\phi}(x,0)} = \ket{{\phi}(x,0)}$.

The drift and diffusion coefficients are obtained as derivatives of
the dispersion relation as $v=\omega_{k_0}^{(1)}$ and
$D=\omega^{(2)}_{k_0} $, where
\begin{equation}
\omega^{(n)}_{k_0} = \left.\frac{\partial^n \omega(k,m)}{\partial k^n}\right|_{k_0}.
\end{equation}
For $|\psi(x,0)\>$ satisfying Eq.~\eqref{eq:smoothstate2},
Eq.~\eqref{eq:dscvs} provides the approximation of the state of the
particle
$|\tilde\psi(x,t)\>=e^{i(k_0x-s\,\omega_0t)}|\tilde\phi(x,t)\>$,
corresponding to
\begin{align} \label{eq:approxstate}
  \ket{\tilde{\psi}(t)} = \int_{-\pi}^{\pi}\,\df{k}{\sqrt{2\pi}} 
e^{-is(\omega_0+vk -\frac12 D k^2 )t }g(k,0) \ket{s}_k \ket{k}.
\end{align} 

The accuracy of this approximation can be quantified in terms of the
parameters $\sigma$ and $\epsilon$ of the initial state
by evaluating (see \ref{a:accuracy}) the overlap between the states
\eqref{eq:smoothstatetimet} and \eqref{eq:approxstate} 
\begin{equation}
|\braket{\tilde{\psi}(t)}{{\psi}(t)}|\geq 1 - \epsilon -\gamma\sigma^3 t - {\cal
  O}(\sigma^5)t,\qquad 
  \gamma= \frac{ \omega^{(3)}_{k_0}}{2 \pi} \int_{k_0-\sigma}^{k_0+\sigma} \d{k}  |g(k,0)|^2.\label{eq:accuracy}
\end{equation}

\begin{figure}[ht!]
  \centering
\includegraphics[width=0.40\textwidth]{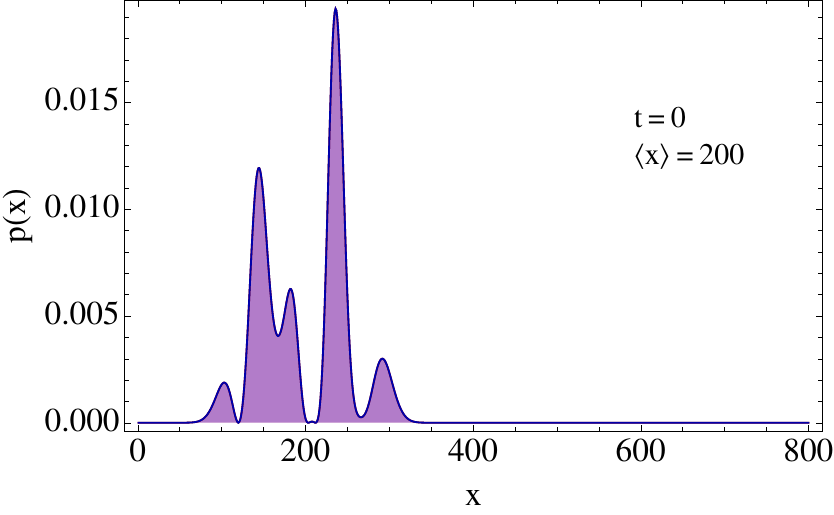}
\qquad \qquad 
\includegraphics[width=0.40\textwidth]{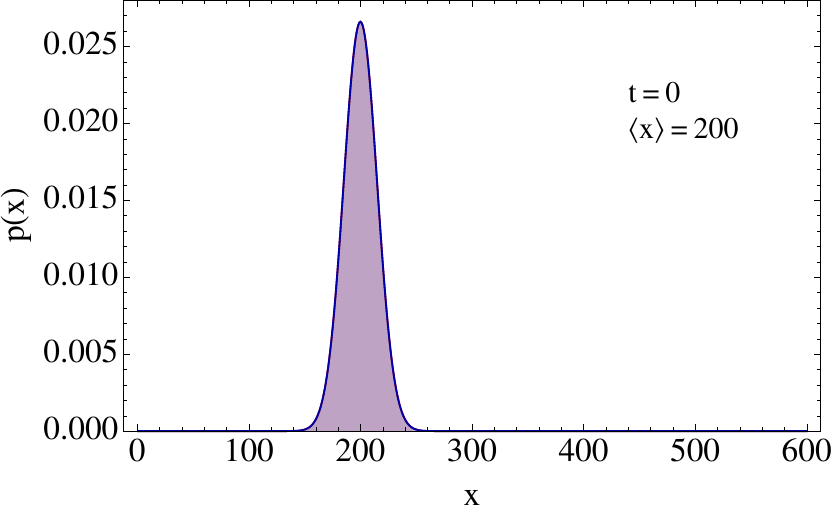}\\
\includegraphics[width=0.40\textwidth]{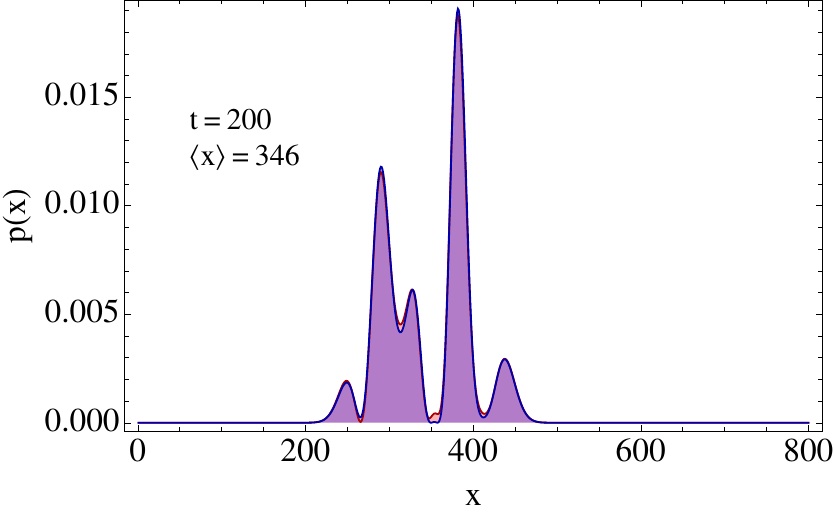}
\qquad \qquad 
\includegraphics[width=0.40\textwidth]{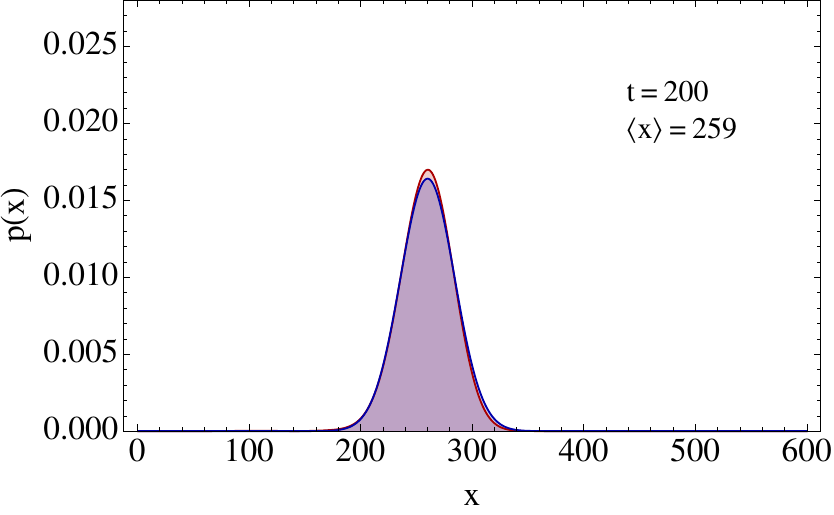}\\
\includegraphics[width=0.40\textwidth]{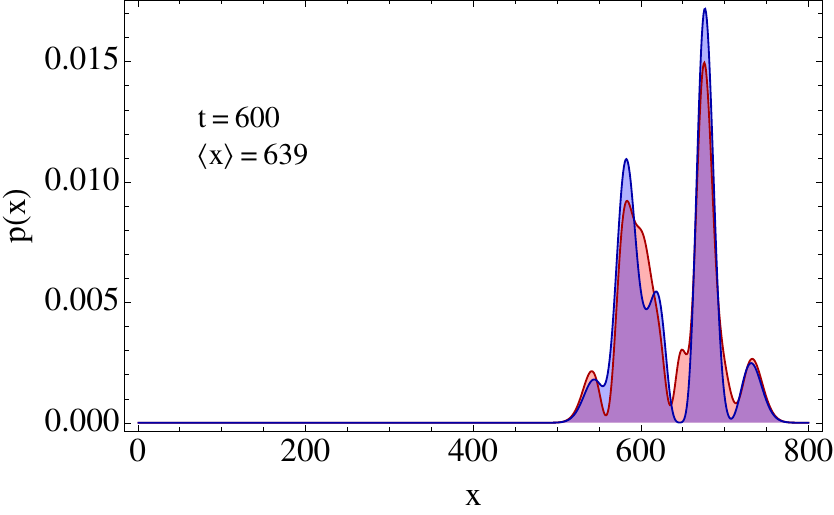}
\qquad \qquad 
\includegraphics[width=0.40\textwidth]{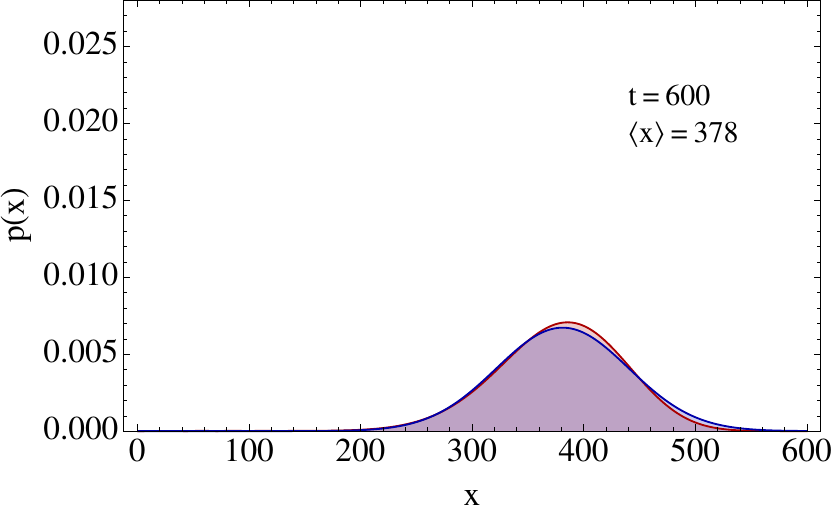}
\caption{(Colors online) Test of the approximation
  \eqref{eq:approxstate} of the Dirac automaton evolution of
  Eq.~\eqref{eq:U} in one space dimension. {\bf Left figure:} here the
  state \eqref{eq:smoothstate} is a superposition of Hermite functions
  (the polynomials $H_j(x)$ multiplied by the Gaussian) peaked around
  momentum $k_0=3\pi/10$, specifically
  $\ket{\v{\psi}(x,0)}=\mathcal{A}e^{ik_0 x}
  \sum_{j\in\mathbb{N}}c_je^{-x^2/4\hat{\sigma}^2}H_j(x/2\hat{\sigma}){\ket{+}_k}_0$
  where $\hat{\sigma}=\sigma^{-1}=20$ is the position variance
  corresponding to the momentum variance $\sigma$, and the
  nonvanishing terms are $c_{0}=\sqrt{1/3}$, $c_2=\sqrt{4/9}$,
  $c_7=\sqrt{2/9}$.  The automaton mass is $m=0.6$. The momentum and
  mass parameters are in the Planckian ultrarelativistic regime. In
  the picture we show a comparison at three different times $t=0$,
  $t=200$ and $t=600$ between the automaton probability distribution
  $|\psi(x,t)|^2$ (in red) and the solution of the differential
  equation (\ref{eq:dscvs}) $|\tilde\psi(x,t)|^2$ (in blue). The drift
  and diffusion coefficients are respectively $v=0.73$ and
  $D=0.31$. The mean position moves at the group velocity given by the
  drift coefficient $v$. The approximation remains accurate even for
  position spread $\hat\sigma=20$ Planck lengths. According to Eq.~\eqref{eq:accuracy} one has significant deviations for $t\approx
  \gamma{\sigma}^3$, which is $t=600$ in the present case. However, a
  reasonable spread $\hat\sigma$ in a typical particle physics
  scenario is the Fermi length $\hat\sigma\approx 10^{20}$, that would
  need a time $t$ comparable to many universe life-times to introduce
  a significant error. The $\epsilon$ error in Eq.~\eqref{eq:accuracy}
  can be taken very small by considering $n\sigma$ instead of $\sigma$
  in Eq.~\eqref{eq:smoothstate2}. For Gaussian states it is enough to
  consider $3\sigma$ to get $\epsilon\approx 10^{-3}$. {\bf Right
    figure:} The same three time comparison for the automaton $m=0.4$,
  and an initial Gaussian state having width
  $\hat\sigma=\sigma^{-1}=10$ and peaked around the momentum
  $k_0=0.1$. In this case the drift velocity and the diffusion
  coefficient are respectively $v=0.22$ and
  $D=2.30$.}\label{fig:Hermite}
\end{figure}

We can test the accuracy of the approximation by comparing it with
the automaton simulation. In Fig.~\ref{fig:Hermite} we show an
example where the initial state \eqref{eq:smoothstate}
is a superposition of Hermite functions (the polynomials $H_j(x)$ multiplied by the Gaussian) peaked
around a very high momentum $k_0=3\pi/10$ and for inertial mass $m=0.6$. The mean value moves at the
group velocity given by the drift coefficient $v$. One can notice how the approximation remains
accurate even for small position spreads of few Planck lengths. For a spread $\hat\sigma$ of the
order of a Fermi as in a typical particle physics scenario, the time $t$ needed for a significant
departure would be comparable to many universe life-times.

In the relativistic regime $k,m\ll 1$ and $k/m\gg 1$, the
dispersive differential equation \eqref{eq:dscvs} approaches the
Dirac equation. The leading order and the corrections to the drift and
diffusion coefficients introduced by the automaton evolution are
\begin{align}
  v=\frac{k}{\sqrt{k^2+m^2}}\label{eq:drift-expansion}
  \left(1-\frac{1}{3}m^2+\frac{1}{6}\frac{m^2k^2}{k^2+m^2}\right),
  \qquad
  D=\frac{m^2}{\sqrt{(k^2+m^2)^3}}\left(1+
    \frac{1}{3}m^2k^2-\frac{1}{2}\frac{m^2k^4}{k^2+m^2}\right).
\end{align}
The leading order in $v$ and $D$ correspond to the Dirac equation.

In the non relativistic regime, $k,m \ll 1$ and $k/m\ll 1$ the usual
Schr\"odinger drift and diffusion coefficients are recovered with the
following corrections
\begin{align}\label{eq:non-rel}
  &v=\frac{k}{m} \left(1+\frac{1}{3}m^2\right),\quad D=\frac{1}{m}\left(1+\frac{5}{6}k^2\right).
\end{align}
Notice that the leading terms are just the usual group-velocity and
diffusion coefficient of the Schr\"odinger equation.

The momentum dependent differential equation (\ref{eq:dscvs}) along with
the leading terms in the relativistic and non relativistic regimes
(see Eq.~\eqref{eq:drift-expansion} and
Eq.~\eqref{eq:non-rel}) provides a useful analytic tool for evaluating the
macroscopic evolution of the automaton, which otherwise would not be
computable in practice.
We now consider an elementary discrimination experiment
between the Dirac automaton evolution and the usual Dirac one based on
particle fly-time. 

Consider again a proton UHECR with $m_p\approx 10^{-19}$ and momentum
peaked around $k_{CR}\approx 10^{-8}$ in Planck units, with a spread
$\sigma$. We ask what is the minimal time $t_{CR}$ for observing a
complete spatial separation between the trajectory predicted by the
cellular automaton model and the one described by the usual Dirac
equation. Thus we require the separation between the two trajectories
to be greater than $\hat\sigma=\sigma^{-1}$ the initial
proton's width in the position space. Notice that UHECR belong
to the relativistic regime $m_p,k_{CR}\ll 1$, where the automaton well approximates the usual
Dirac evolution. We describe the state evolution of the wave-packet of the proton using
the differential equation (\ref{eq:dscvs}) for an initial Gaussian
state. The Dirac evolution corresponds to the 
differential equation (\ref{eq:dscvs}) with drift and diffusion coefficients given by the leading-order
terms in Eq.~\eqref{eq:drift-expansion}, whereas the automaton is described by the full expansion. 
Taking the difference between the drift coefficient in the two cases one can evaluate the time required to have a
separation $\hat\sigma$ between the automaton and the Dirac particle
\begin{align}\label{eq:flying-time1}
  t\approx \hat\sigma\left|\frac{6\sqrt{(k^2+m^2)^3}}{m^2k^2(2m^2+k)}\right|,
\end{align}
that, since it is $m_p/k_{CR}\ll 1$, further simplifies as follows
\begin{align}\label{eq:flying-time2}
t_{CR}\approx 6\frac{\hat\sigma}{m_p^2}.
\end{align}
Furthermore, if we want the separation $\hat\sigma$ to be visible, the broadening $\hat\sigma_{br}(t)$ of
the two packets must be much smaller than $\hat\sigma$. Using Eq.~\eqref{eq:drift-expansion} one
has
\begin{align}
  \begin{split}\nonumber
    \hat\sigma_{br}(t)=\hat\sigma\left(\sqrt{1+\left(\frac{D}{2\hat\sigma^2}t\right)^2}+
      \sqrt{1+\left(\frac{D_{\rm D}}{2\hat\sigma^2}t\right)^2}-2\right)\approx
    2\hat\sigma\left(\sqrt{1+\frac{m_p^4}{4\hat\sigma^4k_{CR}^6}t^2}-1\right)
  \end{split}
\end{align}
where $D_{{\rm D}}=m^2(k^2+m^2)^{-3/2}$ and we used $m_p/k_{CR}\ll 1$. From
Eq.~\eqref{eq:flying-time2} we see that $\hat\sigma\gg\hat\sigma_{br}$ when
\begin{align}
\hat\sigma\gg (k_{CR})^{-3}=10^{22}\;\text{Planck lengths}=10^2{\rm fm}.
\end{align}
With $\hat\sigma=10^2\text{fm}$ (that is reasonable for a proton wave-packet) the flying time request for complete
separation between the two trajectories is
\begin{align}
t_{CR}\approx 6\times 10^{60}\;\text{Planck times}\;\approx 10^{17}s, 
\end{align}
that is comparable with the age of the universe and then incompatible
with a realistic setup. We notice that UHECR, despite being very
energetic, are very rare events and it is not possible to design
experiments involving more that one cosmic ray. Alternatively one
could consider experiments involving many less energetic particles,
reducing the minimal time for the discrimination according to the
theoretical optimal result of Eq.~\eqref{eq:time-scale}, or
experiments based on quantum interferometry and/or ultra-cold atoms as
in
Refs.~\cite{Moyer:2012ws,Hogan:2012ik,pikovski2011probing,amelino2009constraining}.

\section{Conclusions}

In this paper we have considered the evolution of a quantum field in one dimension
via a QCA. The automaton provides the one-step evolution of the 
fields located at the sites $x\in\Z$ of the lattice, inducing a
discrete causal network of points $(x,t)$. The Dirac automaton proposed in Ref.~\cite{darianopla} 
is here derived as the minimum-dimension QCA holding the symmetries of the
causal network, namely the parity and the time reversal invariance. The present one dimensional 
automaton is different from the \emph{coined-quantum walk}, also known as generalized \emph{Hadamard walk}, 
which is usually considered in the QWs literature.

The Dirac automaton, which depends on one parameter $m\in[0,1]$ and has a band-limited wave-vector space $k\in[0,\pi]$,  
is shown to recover the Dirac equation in the limit of small $k$ and $m$, which are then interpreted as the momentum and the
mass of the Dirac field. We proved this result
by considering the problem of discriminating between the Dirac QCA and 
 the usual Dirac evolution for initial states with limited
 momentum and number of particles. We derived an  exact lower bound for the
 probability of error in the discrimination, which is an explicit function of the
mass of the field, the number and the momentum of the particles,
and the duration of the evolution. We observe that for values of these parameters 
compatible with current experiments of particle physics the
probability of error approaches $1/2$
(i.e. the two evolutions are indistinguishable).
We stress that this analysis has not been obtained by taking the continuous limit of the lattice, namely taking
the limit of a sequence of automata with smaller and smaller lattice spacing.

We have then derived an analytical approximation of
the automaton evolution in terms of a dispersive differential
equation. The approximation works for quantum states smoothly peaked around 
some momentum eigenvectors of the 
automaton with the drift and the diffusion coefficients corresponding to the usual Dirac ones 
for small masses and momenta, in accordance to the above rigorous Dirac limit of the automaton.

In the paper \cite{PhysRevA.90.062106}, which is subsequent to the
present one, the derivation of the Dirac QCA has been developed in the 
$2+1$ and in the $3+1$ dimensional cases. One could extend the analysis of this paper to the 
automata of Ref.~\cite{PhysRevA.90.062106} considering the discrimination with their usual Dirac counterparts 
and evaluating of the corresponding dispersive differential equation.

Up to now we have only considered the free field evolution. However, the physical 
interpretation of the automaton dispersion relation and wave-vector as
energy and momentum needs the development of an interacting model. 
Moreover, as we stressed in the introduction, the analysis of this paper
considers a fixed reference frame 
and a major point of the forthcoming research will be the study of relative reference frames within the QCA framework 
and of the analysis of the emerging notion of spacetime.

\section{Acknowledgments}
This work has been supported in part by the Templeton Foundation under
the project ID\# 43796 {\em A Quantum-Digital Universe}.  
We thank Paolo Perinotti for interesting discussions and an anonymous
referee for his/her valuable suggestions.  A. Bisio
and A. Tosini acknowledge useful discussion with Daniel Reitzner.

\appendix

 \section{Derivation of the Dirac automaton}\label{a:derivation}
 In this appendix we present in detail the derivation of the
 Dirac automaton \eqref{eq:U}, here reported in the momentum
 representation (see also Eq.~\eqref{eq:automaton-Uk}) 
\begin{align}\label{eq:U_a}
&\v{U}(k)=\sum_{x\in\{\minus 1,0,1\}} \v{U}_{x}e^{-ikx},\\
&\v{U}_1=\begin{pmatrix}
n&0\\0&0
\end{pmatrix},\quad
\v{U}_{\minus 1}=\begin{pmatrix}
0&0\\0&n
\end{pmatrix},\quad
\v{U}_0=\begin{pmatrix}
0&im\\im&0
\end{pmatrix},\quad n,m\in\Reals^+,\quad n^2+m^2=1,
\end{align}
starting from the assumptions (\ref{1}-\ref{minimal}) of Section
\ref{s:Dirac}. According to assumption \eqref{minimal} we show that
for the internal dimension $\Lambda=1$ there are not admissible non
trivial (non-identical) automata. Then we show that for $\Lambda=2$ there exist non
trivial solutions and that they are all unitarily equivalent to the
one given in Eq.~\eqref{eq:U}. Here, for convenience of the reader, we
report the unitarity conditions (See Eq.~\eqref{eq:unitarity})
\begin{align}
\label{eq:unit1}
  \v{U}_1\v{U}_1^\dag+\v{U}_{\minus 1}\v{U}_{\minus
  1}^\dag+\v{U}_0\v{U}_0^\dag=I,\\\label{eq:unit2}
 \v{U}_0\v{U}_1^\dag+\v{U}_{\minus 1}\v{U}_0^\dag=0,\\\label{eq:unit3} 
\v{U}_{\minus 1}\v{U}_1^\dag=0,
\end{align}
and the parity and time reversal covariance condition (see Eqs.~\eqref{eq:parity}
and \eqref{eq:time-reversal})
\begin{align}\label{eq:parity_a}
\v{P}\v{U}_{\pm1}\v{P}^\dag=\v{U}_{\mp 1},\quad \v{P}\v{U}_0\v{P}^\dag=\v{U}_0,\\\label{eq:time-reversal_a}
\v{T}\v{U}_{\pm 1}\v{T}^\dag=\v{U}_{\mp 1}^\dag,\quad \v{T}\v{U}_0\v{T}^\dag=\v{U}_0^\dag, 
\end{align}
for some unitary $\v{P}$ and anti-unitary $\v{T}$ operators
\footnote{Any anti-unitary operator $\v{T}$ is given by $\v{T} =
  \v{C}\v{U}$, where $\v{U}$ is a unitary operator and $\v{C}$ is the
  complex \emph{conjugation operator} (given a basis
  $\{\ket{\alpha_i}\}$ of a Hilbert space $\Hilb$ and an arbitrary
  vector $\ket{\alpha}=\sum c_i\ket{\alpha_i}$ it is $\v{C}(\sum_i
  c_i\ket{\alpha_i})=\sum_i c^*_i\ket{\alpha_i}$). Here we briefly
  recall the reason why the time reversal symmetry $\v{T}$,
  interchanging the forward and backward light-cones $(t,x)\rightarrow
  (-t,x)$, cannot be represented by a unitary but by an anti-unitary
  operator. Take an eigenstate of the automaton $\ket{s}_k$, with
  $\v{U}(k) \ket{s}_k=e^{i\v{H}(k)}\ket{s}_k=e^{-is\omega}\ket{s}_k$,
  and consider the two states $\ket{\psi}_1=\v{T}e^{-is\omega
    t}\ket{s}_k $ and $\ket{\psi}_2=e^{is\omega t}\v{T}\ket{s}_k $. In
  the first case the state is evolved forward in time and then the
  time reversal is applied, in the second case we first act with the
  time-reversal operator and then evolve backward in time the
  state. If $\v{T}$ is a symmetry of the Dirac theory the two
  operations must commute and one gets
  $\ket{\psi}_1=\ket{\psi}_2\Rightarrow \v{T}e^{-is\omega
    t}\ket{s}_k =e^{is\omega t}\v{T}\ket{s}_k $, which shows the
  non-linear action of the $\v{T}$ operator.}.

\subsection{$\Lambda=1$}
For $\Lambda=1$, the transition matrices are just complex numbers, say
\begin{align}
\v{U}_{1}=e^{i\theta},\quad\v{U}_{\minus 1}=e^{i\theta'},\quad\v{U}_0=e^{i\theta''}.
\end{align}
In this case the unitarity constraints
Eqs.~\eqref{eq:unit1},\eqref{eq:unit2}, and \eqref{eq:unit3} lead to
only three possible solutions
\begin{align}
 \v{U}_{1}=e^{i\theta}, \v{U}_0=\v{U}_{\minus 1}=0,\qquad
 \v{U}_{\minus 1}=e^{i\theta}, \v{U}_0=\v{U}_{1}=0,\qquad
 \v{U}_0=e^{i\theta},  \v{U}_{1}=\v{U}_{\minus 1}=0,
\end{align}
with $\theta\in[0,2\pi]$. Modulo a global phase, the above solutions
correspond respectively to the right-shift ($\v{U}=S_1$), the
left-shift ($\v{U}=S_{\minus 1}$) and the identical ($\v{U}=I$)
automaton. Since in the right- and the left-shift solutions only one of
the two transition matrices $\v{U}_1$, $\v{U}_{\minus 1}$ is not
null, parity covariance \eqref{eq:time-reversal_a} cannot be satisfied
and we are left with the trivial solution corresponding to the
identical automaton.

\subsection{$\Lambda=2$}
For $\Lambda=2$ the three transition matrices can be generally
parametrized as follows
\begin{align}
\label{eq:a1}
\v{U}_1=\begin{pmatrix}
a&b\\c&d
\end{pmatrix},\quad
\v{U}_{\minus 1}=\begin{pmatrix}
a'&b'\\c'&d'
\end{pmatrix},\quad
\v{U}_0=\begin{pmatrix}
x&y\\z&w
\end{pmatrix},
\end{align}
with all entries arbitrary complex numbers. 

Now we can fix the basis where $\v{P}$ and $\v{T}$ in
Eqs.~\eqref{eq:parity_a} and \eqref{eq:time-reversal_a} are
represented as
\begin{align}
\label{eq:a2}
\v{P}=\begin{pmatrix}
0&1\\1&0
\end{pmatrix},\qquad
\v{T}=\v{C}\begin{pmatrix}
0&1\\1&0
\end{pmatrix},
\end{align}
where $\v{C}$ is the anti-unitary operator denoting complex
conjugation in the given representation.  Indeed without loss of
generality we can fix the representation (which fix a basis) for one
of the two symmetries, say parity.  Once parity is given we have to
represent time reversal in the same basis, with different choices
leading in general to non unitary equivalent solutions. However,
assuming $[\v{P},\v{T}]=0$ as it is in the usual QFT (we do not
consider the more general scenario where the two operators do not
commute), and discarding the trivial case where $\v{T}\propto I$, we
are left with the representation of Eq.~\eqref{eq:a2}.

In the representation \eqref{eq:a2} the parity covariance
\eqref{eq:parity_a} of the automaton gives
\begin{align}\label{eq:a3}
\v{U}_1=\begin{pmatrix}
a&b\\c&d
\end{pmatrix},\quad
\v{U}_{\minus 1}=\begin{pmatrix}
d&c\\b&a
\end{pmatrix},\quad
\v{U}_0=\begin{pmatrix}
x&y\\y&x
\end{pmatrix},
\end{align}
while from the time time-reversal covariance
\eqref{eq:time-reversal_a} it follows
\begin{align}\label{eq:a4}
\v{U}_1=\begin{pmatrix}
a&b\\b&d
\end{pmatrix},\quad
\v{U}_{\minus 1}=\begin{pmatrix}
d&b\\b&a
\end{pmatrix},\quad
\v{U}_0=\begin{pmatrix}
x&y\\y&x
\end{pmatrix}.
\end{align}

Equation \eqref{eq:parity_a} shows that $\v{U}_1$ and $\v{U}_{\minus 1}$
are unitarily equivalent (they are related by conjugation with the
unitary operator $\v{P}$), and from the condition $\v{U}_{\minus
  1}\v{U}_1^\dag=0$ in \eqref{eq:unit3} it follows that they are
both rank one.  Accordingly, without loss of generality, we can always
write the two transition matrices as follows
\begin{align}\label{eq:a5}
\v{U}_1=\begin{pmatrix}
a&b\\\eta a&\eta b
\end{pmatrix},\quad
\v{U}_{\minus 1}=\begin{pmatrix}
\eta b&\eta a\\\ b& a
\end{pmatrix},
\end{align}
for some $\eta\in\mathbb{C}$.
Now we consider separately the two cases $\eta= 0$, and $\eta\neq 0$.

($\eta=0$) From the time reversal invariance
\eqref{eq:time-reversal}, more precisely from
$\v{T}\v{U}_1\v{T}^\dag=\v{U}_{\minus 1}^\dag$, it follows
$b=0$. Using this result the unitarity condition \eqref{eq:unit2}
gives the two equalities $xa^*=ax^*=0$ and $ya^*+ay^*=0$. Since the
case $a=0$ is trivial ($\v{U}_{1}=\v{U}_{\minus 1}=0$) it follows
$x=0$ and $\Re(ay^*)=0$. Finally, using the unitarity condition
\eqref{eq:unitarity} we get $|a|^2+|y|^2=1$ that, up to a global
phase, gives the unique solution
\begin{align}\label{eq:aU}
\v{U}(k)=\begin{pmatrix}
  n e^{ik} &-im\\
  -im & ne^{-ik}
\end{pmatrix},\quad n,m\in\Reals,\quad n^2+m^2=1\,.
\end{align}
The constants $n$ and $m$ in the last equation can be chosen positive
since a change in the relative sign is obtained by a unitary
conjugation with the matrix
$\left(\begin{smallmatrix}0&-i\\i&0\end{smallmatrix}\right)$.

($\eta\neq 0$)
Noticing that for Eq.~\eqref{eq:a4} it must  be $\eta a= b$ we have
\begin{align}\label{eq:a6}
\v{U}_1=\begin{pmatrix}
b/\eta&b\\b&\eta b
\end{pmatrix},\quad
\v{U}_{\minus 1}=\begin{pmatrix}
\eta b& b \\b&b/\eta
\end{pmatrix},
\end{align}
and using again the condition $\v{U}_{\minus 1}\v{U}_1^\dag=0$ in
\eqref{eq:unitarity}, with $\v{U}_{\pm 1}$ as in Eq.~\eqref{eq:a6} , we
get the constraints
\begin{align}\label{eq:7}
|b|^2(\eta/\eta^*+1)=|b|^2(\eta+\eta^*)=0.
\end{align}
Since the case $b=0$ is trivial, we take $b\neq 0$ in which case
\eqref{eq:7} implies $\Re(\eta)=0$, say
\begin{align}\label{eq:a5}
\v{U}_1=\begin{pmatrix}
-ib/\xi &b\\ b&i \xi b
\end{pmatrix},\quad
\v{U}_{\minus 1}=\begin{pmatrix}
i\xi b &b\\ b&-i b /\xi
\end{pmatrix},
\end{align}
for some $\xi\in\Reals$ and with $\xi\neq 0$. Using the
unitarity conditions \eqref{eq:unit1} and \eqref{eq:unit2} we get
respectively the equalities
\begin{align}\label{eq:norm}
& |x|^2+|y|^2+|b|^2(1+\xi^2+1/\xi^2)=1,\\\label{eq:xy}
& xy^*+yx^*=0,
\end{align}
and
\begin{align}\label{eq:yb1}
&yb^*-by^*=yb^*+by^*=0\\
&xb^*+bx^* =xb^*+\xi^2bx^* =0
\end{align}
Since Eq.~\eqref{eq:yb1} implies both $y=pb$ and $y=iqb$ for some
$p,q\in\Reals$, and $b\neq 0$ by hypothesis, it must be $y=0$. Moreover, due to
Eq.~\eqref{eq:xy} which gives $x=iry$ for some $r\in\Reals$, we get $x=0$
proving that the transition matrix $\v{U}_0$ is the null matrix.  
Using Eq.~\eqref{eq:norm} we find
\begin{align}
\frac{(\xi^2+1)^2}{\xi^2}|b|^2=1\Rightarrow b=e^{i\theta}\frac{\xi}{\xi^2+1}.
\end{align}
with $\theta\in\Reals$ and the general solution for $\eta\neq 0$, up to a global phase, is finally given by
\begin{align}\label{eq:sol2}
\v{U}(k)=
\begin{pmatrix}
i\frac{\xi^2}{\xi^2+1}e^{ik}-i\frac{1}{\xi^2+1}e^{-ik}&\frac{\xi}{\xi^2+1}(e^{ik}+e^{-ik})\\
\frac{\xi}{\xi^2+1}(e^{ik}+e^{-ik})&i\frac{\xi^2}{\xi^2+1}e^{-ik}-i\frac{1}{\xi^2+1}e^{ik}
\end{pmatrix}.
\end{align}

Now we observe that the dispersion relation of the solutions
\eqref{eq:aU} and Eq.~\eqref{eq:sol2}, corresponding respectively to
the cases $\eta=0$ and $\eta\neq 0$, are given by
\begin{align}
\omega_{\eta=0}(k)=\arccos{(n\cos{(k)})},\qquad
\omega_{\eta\neq 0}(k)=\arccos{\left(\tfrac{\xi^2-1}{\xi^2+1}\cos{(k)}\right)},
\end{align}
which coincide upon the identification $n=\tfrac{\xi^2-1}{\xi^2+1}$
(this is always possible because both $n$ and
$\tfrac{\xi^2-1}{\xi^2+1}$ are real numbers smaller or equal to
one). Since the automata in Eq.~\eqref{eq:aU} and Eq.~\eqref{eq:sol2}
have the same dispersion relation they have the same eigenvalues
$e^{\pm i\omega}$ are then unitarily equivalent.

\section{Proof of the bound (\ref{eq:ilboundperpe})} 
\label{a:large-scale}
In this appendix we detail the proof of the bound
\eqref{eq:ilboundperpe} in Section \ref{s:DiracB}
which provides the probability of optimal error probability in
discriminating the Dirac automaton and the usual Dirac evolution.  The
discrimination experiment can have a generic duration $t$ and the
unitary operators to be discriminated are explicitly given by

\begin{align}
\label{eq:unitaryautom}
\begin{split}
\v{U}^t(k)=\exp({-i\v{H}(k)t})=\begin{pmatrix} \cos(\omega t) + i \frac{\sin(\omega t)}{\omega} a &
  - i b
  \frac{\sin(\omega t)}{\omega} \\
  - i b \frac{\sin(\omega t)}{\omega} &
  \cos(\omega t) - i \frac{\sin(\omega t)}{\omega} a \\
\end{pmatrix}
\\
a := \frac {\omega}{\sin(\omega)} n \sin(k) \qquad b := \frac
{\omega}{\sin(\omega)} m
\end{split}
\end{align}

\begin{align}
\label{eq:unitarydirac}
\begin{split}
\v{U}^t_{\rm D}(k)=\exp{(-i\v{H}_{\rm D}(k)t})=\begin{pmatrix} \cos(\lambda t) + i \frac{\sin(\lambda t)}{\lambda} k
  & - i m
  \frac{\sin(\lambda t)}{\lambda} \\
  - i m \frac{\sin(\lambda t)}{\lambda} &
  \cos(\lambda t) - i \frac{\sin(\lambda t)}{\lambda} k \\
  \end{pmatrix},
\end{split}
\end{align}
as can be easily verified by direct computation using the Hamiltonians
in Eqs. \eqref{eq:Hunp} and \eqref{eq:dirac-hamiltonian}. The
proof of the bound \eqref{eq:ilboundperpe} goes through the following
three Lemmas.

\begin{lemma}\label{l:bound1}
Let $\v{U}^t_{\rm D}(k)$ and $\v{U}^{t}(k)$ be defined according to
Eqs. \eqref{eq:unitaryautom},\eqref{eq:unitarydirac} and let us define
$\v{V}(k,t) = \v{U}^t_{\rm D}(k) \v{U}^{t\dagger}(k)$. Let $e^{ i
  \mu(k,m,t)}$ be an eigenvalue of $\v{V}(k,t)$.  Then the following
bound holds:
  \begin{align}
    \label{eq:12}
    \cos(\mu(k,m,t)) \geq \cos(\alpha t) -\beta
  \end{align}
  where
  \begin{align}
    \begin{split}
    \alpha(k,m) := \omega_{\rm D} - \omega& \\
    \beta(k,m) 
:= \frac12 \left( 1 - vv_{\rm D} -
      \sqrt{(1-v^2)(1-v_{\rm D}^2)} \right)&\,. 
  \end{split}\label{eq:alphabeta}
\end{align}
\end{lemma}
\Proof
Since both 
$\v{U}^t_{\rm D}(k)$ and $\v{U}^{t\dagger}(k)$
are $SU(2)$ matrices, we have that 
$\v{V}(k,t)$ is an $SU(2)$ matrix
and its eigenvalues must be of the form 
$e^{ i \mu(k,m,t)}$ and $e^{- i \mu(k,m,t)}$.
This implies the equality
$  \cos(\mu(k,m,t)) = \frac12 \Tr[\v{V}(k,t)] $
which by direct computation gives
\begin{align}
   \cos(\mu(k,m,t)) = \left(1-\frac{\beta}{2} \right)\cos(\alpha t) +
 \frac{\beta}{2} \cos(\gamma t) 
\label{eq:cosenomu}
\end{align}
where $\alpha$ and $\beta$ are defined accordingly with Eq.~\eqref{eq:alphabeta} and $\gamma := \omega + \omega_{\rm D}$.
Finally, from Eq.~\eqref{eq:cosenomu} one has the bound
$\cos(\mu(k,m,t)) \geq \cos(\alpha t) -\beta$ \qed

The second  Lemma shows the monotonicity of the two functions
$\alpha,\beta$ in Lemma \ref{l:bound1}:
\begin{lemma}
  Let $\alpha(k,m)$ and
$\beta(k,m)$ be defined as in Eq.~\eqref{eq:alphabeta}
and $0 \leq \bar{k} < \pi$
Then we have
\begin{align}
  \begin{split}
    \bar{\alpha}:= \max_{k \in [-\bar{k}, \bar{k}]} |\alpha| =
    \max_{k \in \{0, \bar{k}\}} |\alpha|\\
    \bar{\beta}:= \max_{k \in [-\bar{k}, \bar{k}]} |\beta| =
    \max_{k \in \{0, \bar{k}\}} |\beta|
  \end{split}
\qquad\forall m \in [0,1]\,. \label{eq:alphabetagrow}
\end{align} 
\end{lemma}

\Proof Since both $\omega$ and $\omega_{\rm D}$ are even functions of
$k$, from Eq.~\eqref{eq:alphabeta} we have that also $\alpha$ and
$\beta$ are even function of $k$.  For this reason we can restrict to
$k\in[0,\bar{k}] $.  The equality \eqref{eq:alphabetagrow} can be
proved by showing that $\alpha$ and $\beta$ are nondecreasing functions
of $k$ for $k\in[0,\bar{k}] $.

Since $\partial_k\alpha=v_{\rm D}-v$, clearly $v_{\rm D}^2 -v^2\geq 0$ for
$k\in[0,\pi)$ implies $\partial_k\alpha \geq 0$ in the same interval.
By direct computation one can verify that
\begin{align}
  (v_{\rm D})^2 -(v)^2 = \frac{x(k,m)} {y(k,m)}& \\
x(k,m) := k^2- \sin^2(k)(1-m^2)&\\
y(k,m):= (k^2 +m^2)(\sin^2(k) + m^2 \cos^2(k))&\,.
\end{align}
Clearly we have $y(k,m) \geq 0$ and since $k \geq \sin(k)$ for $0 \leq
k < \pi$, the thesis is proved. 

Again the monotonicity of $ \beta$ for $k \in [0,\pi )$ follows from
$\partial_k\beta\geq 0$ in the same interval. By elementary
computation we have
\begin{align}
  \partial_k\beta=x(k,m)y(k,m)z(k,m)\\
x(k,m):=\frac{m^2}{\omega_{\rm D}\sin^2(\omega)}\\
y(k,m):=(n\sin(k)-k)\\
z(k,m):=\frac{n\cos(k)}{\sin^2(\omega)}-\frac{1}{\omega_{\rm D}^2}
\end{align}
Clearly $x(k,m)y(k,m)\leq 0$ for $k\in[0,\pi)$ and we just have to
verify that $z(k,m)\leq 0$ in that interval, namely 
\begin{align}\label{eq:dis1} 
m^2\cos^2(k)+\sin^2(k)-n \cos(k)\omega_{\rm D}^2\geq 0\,. 
\end{align}
The last equation is clearly satisfied for $k\in[\pi/2,\pi]$ therefore
we restrict to $k\in[0,\pi/2]$. This allows to divide the left side of
Eq.~\eqref{eq:dis1} by $\cos(k)$ achieving
\begin{align}
m^2\cos(k)+\frac{\sin^2(k)}{\cos(k)}-n\omega_{\rm D}^2\geq 0 
\end{align}
which is satisfied if
\begin{align}
w(k,m):=m^2\cos(k)+\sin^2(k)-n\omega_{\rm D}^2 \geq 0\,.
\end{align}
It is easy to see that, for any $m\in[0,1]$, we have
$\left(\partial^{(i)}_kw(k,m)\right)_{k=0}=0$ for $i=0,1$, while
$\partial^{(2)}_kf(k,m)\geq 0$ for any $k\in[0,\pi/2]$, which gives
the monotonicity of $\beta$.  \qed 

\begin{lemma}\label{l:cruciale}
  Let $0 \leq \bar{k} < \pi$, $\bar{N}$ be a positive integer number, and
  $\bar{\alpha}$, $\bar{\beta}$ be defined as in
  Eq.~\eqref{eq:alphabetagrow}. If $\bar{\beta} \leq 1- \cos
  (\frac{\pi}{2\bar{N}})$ and $t \leq f(\bar{k},m,\bar{N})$ where
  \begin{align}
    f(\bar{k},m,\bar{N}) := \frac{\arccos(\cos\left( \frac{\pi}{2\bar{N}}\right) +
      \bar{\beta})}{\bar{\alpha}}
 \end{align}
 then
 \begin{align}
  \label{eq:2}
  \bar{N} \mu(k,m,t)) \leq g(\bar{k},m,\bar{N},t) \leq \frac{\pi}{2} 
\end{align}
where $g(\bar{k},m,\bar{N},t):= \bar{N}\arccos\left( \cos({\bar{\alpha}}t)
  -{\bar{\beta}} \right)$.
\end{lemma}

\Proof
The conditions $t \leq f(\bar{k},m,\bar{N})$ and $\bar{\beta} \leq 1-
\cos (\frac{\pi}{2\bar{N}})$
imply
\begin{multline}
0 \leq  \bar{\alpha}t \leq \arccos(\cos\left(
  \frac{\pi}{2\bar{N}}\right) + \bar{\beta})
\Rightarrow
1 \geq \cos(\bar{\alpha}t) -\bar{\beta} \geq \cos\left(
  \frac{\pi}{2\bar{N}}\right)
\Rightarrow \\
\Rightarrow
 \cos({\alpha}t) -{\beta} \geq \cos(\bar{\alpha}t)
-\bar{\beta} \geq\cos\left(
  \frac{\pi}{2\bar{N}}\right)\,. \label{eq:incubocoseni}
\end{multline}
By exploiting the bound \eqref{eq:12} into Eq.~\eqref{eq:incubocoseni}
we have
\begin{multline}
  \cos(\mu(k,m,t)) \geq \cos({\bar{\alpha}}t) -{\bar{\beta}} \geq \cos\left(
  \frac{\pi}{2\bar{N}}\right) \Rightarrow 
\bar{N} \mu(k,m,t) \leq \bar{N}\arccos\left( \cos({\bar{\alpha}}t) -{\bar{\beta}} \right) \leq
\frac{\pi}{2}\,.
\end{multline}
\qed 

We are now ready to prove the bound \eqref{eq:ilboundperpe}
\begin{proposition}
  Let $U^{t}$ and $U_{\rm D}^{t}$ be the unitary evolutions given by the
  Dirac QCA and by the Dirac equation respectively.  If the hypothesis
  of Lemma \ref{l:cruciale} holds we have
\begin{align}
  \label{eq:ilverobound}
  \sup_{\rho \in \mathcal{T}_{\bar{k},\bar{N}}} || (U^{t}\rho U^{t
    \dagger} - U_{\rm D}^{t} \rho U_{\rm D}^{t\dagger}) ||_1 \leq
  \sqrt{1-\cos^2(g(\bar{k},m,\bar{N},t))}.
\end{align}
\end{proposition}

\Proof First we notice that thanks to the convexity of the trace
distance we can without loss of generality consider $\rho$ to be pure.
If $\rho$ is a pure state $\ketbra{\chi}{\chi}$ the trace distance
becomes $ \sqrt{1-|\bra{\chi} U^{t}U_{\rm D}^{t \dagger} \ket{\chi} |^2}=
\sqrt{1-|\bra{\chi} V(t) \ket{\chi} |^2}$.  If we expand $\ket{\chi}$
on a basis of eigenstates of $V$, i.e. $\ket{\chi} =
\sum_{N,\v{k},\v{s}} \sqrt{p_{N,\v{k},\v{s}}} \ket{N,
  \v{k},\v{s}}$, we have
\begin{align}
    |\bra{\chi} V(t) \ket{\chi}  | = 
\left|
\sum_{N,\v{k},\v{s}} p_{N,\v{k},\v{s}}
       \exp \left( i \sum_{j=0}^{{N}} s_j \mu(k_j,m,t) \right)
\right| \geq\left|\sum_{N,\v{k},\v{s}} p_{N,\v{k},\v{s}} \cos\left(\sum_{j=0}^{{N}} s_j\mu(k_j,m,t) \right) 
      \right|.\label{Eq.boundnorm}
\end{align}
By exploiting the bound \eqref{eq:2} into Eq.~\eqref{Eq.boundnorm} we have
\begin{align*}
  \left|\sum_{N,\v{k},\v{s}} p_{N,\v{k},\v{s}}
    \cos\left(\sum_{j=0}^{{N}} s_j\mu(k_j,m,t) \right)
  \right|^2 \geq \cos^2(g(\bar{k},m,\bar{N},t))
\end{align*}
which finally implies
\begin{align*}
  \sqrt{1-|\bra{\chi} V(t) \ket{\chi}  |^2} \leq 
\sqrt{1-\cos^2(g(\bar{k},m,\bar{N},t))}\,.
\end{align*}
\qed 
Inserting the bound \eqref{eq:ilverobound} into
Eq.~\eqref{eq:supoverstate} we finally have the bound
\eqref{eq:ilboundperpe}.

\section{Derivation of Eq.~\eqref{eq:accuracy}}
\label{a:accuracy}
Here we evaluate the overlap between the exact automaton evolution
$\ket{\psi(t)}$ and the dispersive differential equation approximation
$\ket{\tilde\psi(t)}$
\begin{align}
  |\braket{\tilde{\psi}(t)}{{\psi}(t)}| =& \left|
    \int_{-\pi}^{\pi}\,\df{k}{2\pi} e^{-i( \omega^{(3)}_{k_0}k^3 +{\cal O}(k^4) )t }|g(k,0)|^2 \right|
  \nonumber\\
  \geq& \left|\frac{1}{2\pi} \int_{k_0-\sigma}^{k_0+\sigma} \d{k}  e^{-i( \omega^{(3)}_{k_0}k^3 + {\cal O}(k^4) 
      )t }|g(k,0)|^2 \right|
- \left|\frac{1}{2\pi} \int_{ |k-k_0 |\geq \sigma} \!\!\!\!\!
    \!\!\!\!\! \!\!\!\!\! \d{k}  e^{-i( \omega^{(3)}_{k_0}k^3 + {\cal O}(k^4)
      )t }|g(k,0)|^2 \right| 
  \nonumber\\
  \geq&\left| 1 - i t \frac{ \omega^{(3)}_{k_0} \sigma^3}{2 \pi}
    \int_{k_0-\sigma}^{k_0+\sigma} \d{k} 
    |g(k,0)|^2
    - {\cal O}(\sigma^5)t \right|- \epsilon\nonumber\\
  \geq& 1 - \epsilon -\gamma\sigma^3 t - {\cal O}(\sigma^5)t\nonumber
\end{align}
with the constant $ \gamma= \frac{ \omega^{(3)}_{k_0}}{2 \pi}
\int_{k_0-\sigma}^{k_0+\sigma}\, \d{k}  |g(k,0)|^2$.

\bibliographystyle{model1a-num-names}
\bibliography{bibliography}

\end{document}